\pdfoutput=1
\documentclass[11pt, leqno]{article}
\usepackage[total={6.5in,9in}, top=1in, left=1 in]{geometry}

\usepackage{amsmath, amsthm, amssymb}
\usepackage{graphicx}
\usepackage{epsfig,psfrag}
\usepackage{MnSymbol}
\usepackage{setspace}
\usepackage{subfigure}
\usepackage{epstopdf}
\usepackage{algorithmic}
\usepackage{verbatim} 
\usepackage{enumerate}
\usepackage{url}
\usepackage{color}
\usepackage{pdfsync}
\usepackage{times}
\newtheorem{theorem}{Theorem}
\newtheorem{lemma}[theorem]{Lemma}

\newtheorem{corollary}[theorem]{Corollary}

\newtheorem{remark}[theorem]{Remark}

\doublespace

\newcommand{\eps}{\epsilon}
\newcommand{\abo}{[\alpha / \beta]_O}
\newcommand{\abt}{[\alpha / \beta]_T}

\newcommand{\at}{\alpha_T}
\newcommand{\bt}{\beta_T}
\newcommand{\g}{\gamma}

\newcommand{\sumD}{\sum_{k=0}^{N-1} d_k}

\newcommand{\yjp}{Y_{j}^+}
\newcommand{\yjm}{Y_{j}^-}
\newcommand{\yip}{Y_{i}^+}

\begin{document}

\begin{center}
\LARGE
Optimization of Radiation Therapy Fractionation Schedules in the Presence of Tumor Repopulation
\end{center}
\begin{center}
Thomas Bortfeld\footnote{Department of Radiation Oncology, Massachusetts General Hospital and Harvard Medical School, tbortfeld@partners.org.}, Jagdish Ramakrishnan\footnote{Wisconsin Institute for Discovery, University of Wisconsin, jramakrishn2@wisc.edu, This research was performed while this author was with the Laboratory for Information and Decision Systems, Massachusetts Institute of Technology.}, John N. Tsitsiklis\footnote{Laboratory for Information and Decision Systems, Massachusetts Institute of Technology, jnt@mit.edu.}, and Jan Unkelbach\footnote{Department of Radiation Oncology, Massachusetts General Hospital and Harvard Medical School, junkelbach@partners.org.}
\end{center}

\vspace{-10pt}

\begin{abstract}
We analyze the effect of tumor repopulation on optimal dose delivery in radiation therapy. We are primarily motivated by accelerated tumor repopulation towards the end of radiation treatment, which is believed to play a role in treatment failure for some tumor sites. A dynamic programming framework is developed to determine an optimal fractionation scheme based on a model of cell kill due to radiation and tumor growth in between treatment days. We find that faster tumor growth suggests shorter overall treatment duration. In addition, the presence of accelerated repopulation suggests larger dose fractions later in the treatment to compensate for the increased tumor proliferation. We prove that the optimal dose fractions are increasing over time. Numerical simulations indicate potential for improvement in treatment effectiveness.
\end{abstract}

\section{Introduction}

According to the American Cancer Society, at least 50\% of cancer patients undergo radiation therapy over the course of treatment. Radiation therapy plays an important role in curing early stage cancer, preventing metastatic spread to other areas, and treating symptoms of advanced cancer. For many patients, external beam radiation therapy is one of the best options for cancer treatment. Current therapy procedures involve taking a pre-treatment Computed Tomography (CT) scan of the patient, providing a geometrical model of the patient that is used to determine incident radiation beam directions and intensities. In current clinical practice, most radiation treatments are fractionated, i.e., the total radiation dose is split into approximately $30$ fractions that are delivered over a period of $6$ weeks. Fractionation allows normal tissue to repair sublethal radiation damage between fractions, and thereby tolerate a much higher total dose. Currently, the same dose is delivered in all fractions, and temporal dependencies in tumor growth and radiation response are not taken into account. Biologically-based treatment planning, aiming at optimal dose delivery over time, has tremendous potential as more is being understood about tumor repopulation and reoxygenation, healthy tissue repair, and redistribution of cells. 

In this paper, we study the effect of tumor repopulation on optimal fractionation schedules, i.e., on the total number of treatment days and the dose delivered per day. We are primarily motivated by accelerated tumor repopulation towards the end of radiation treatment, which is considered to be an important cause of treatment failure, especially for head and neck tumors (\cite{WTM1988,Wit1993}). Our main conclusion is that accelerated repopulation suggests larger dose fractions later in the treatment to compensate for the increased tumor proliferation.

\subsection{Motivation}

Radiation therapy treatments are typically fractionated (i.e., distributed over a longer period of time) so that normal tissues have time to recover. However, such time in between treatments allows cancer cells to proliferate and can result in treatment failure (\cite{KiT2005}). The problem of interest then is the determination of an optimal fractionation schedule to counter the effects of tumor repopulation. Using the biological effective dose (BED) model, a recent paper (\cite{MTD2012}) has mathematically analyzed the fractionation problem in the absence of repopulation. For a fixed number of treatment days, the result states that the optimal fractionation schedule is to either deliver a single dose or an equal dose at each treatment day. The former schedule of a single dose corresponds to a hypo-fractionated regimen, in which treatments are ideally delivered in as few days as possible. The latter schedule of equal dose per day corresponds to a hyper-fractionation regimen, in which treatments are delivered in as many days as possible. The work in this paper further develops the mathematical framework in \cite{MTD2012} and analyzes the effect of tumor repopulation on optimal fractionation schedules. We are interested in optimizing \emph{non-uniform} (in time) dose schedules, motivated primarily by the phenomenon of accelerated repopulation, i.e., a faster repopulation of surviving tumor cells towards the end of radiation treatment.

\subsection{Related Work}

There has been prior work on the optimization of non-uniform radiation therapy fractionation schedules (\cite{AlB1976,Swa1981,Swa1984,MaD1991,YLH1994,YaX2005}), some of which also includes tumor repopulation effects. However, these works have either not used the BED model or have primarily considered other factors such as tumor re-oxygenation. It has been shown that effects such as re-oxygenation, re-distribution, and sublethal damage repair can result in non-uniform optimal fractionation schemes (\cite{YaX2005,BBP2013}). Previous works have considered the case of exponential tumor growth with a constant rate of repopulation (\cite{WKO1977,JTD1995,ADJ2004}). Other tumor growth models, e.g. Gompertzian and logistic, have also been considered although mostly in the context of constant dose per day (\cite{Ush1980,McO2007}). 

There is a significant amount of literature, especially from the mathematical biology community, on the use of control theory and dynamic programming (DP) for optimal cancer therapy. Several of these works (\cite{ZiN1979,PeV1991,LeS2004,SUB2014}) have looked into optimization of chemotherapy. For radiation therapy fractionation, some studies (\cite{HeW1973,AlB1976,WCW2000}) have used the DP approach based on deterministic biological models, as done in this paper. However, these works have not carried out a detailed mathematical analysis of the implications of optimal dose delivery in the presence of accelerated repopulation. Using imaging information obtained between treatment days, dynamic optimization models have been developed to adaptively compensate for past accumulated errors in dose to the tumor (\cite{FeV2004,ZAX2007,DeF2008,SEP2010}). There also has been work on online approaches which adapt the dose and treatment plan based on images obtained immediately prior to treatment (\cite{LCC2008,CLC2008,KGP2009,Kim2010,Gha2011,KGP2012,RCB2012}).

Perhaps the closest related work is \cite{WCW2000}, which considers both faster tumor proliferation and re-oxygenation during the course of treatment. While a dose intensification strategy is also suggested in \cite{WCW2000}, the primary rationale for increasing dose fractions is different: it is concluded that due to the increase in tumor sensitivity from re-oxygenation, larger fraction sizes are more effective at the end of treatment. Our work, on the other hand, suggests dose intensification (i.e., larger doses over time) as a direct consequence of a model of accelerated tumor repopulation during the course of treatment.

\subsection{Overview of Main Contributions} \label{m1}

The primary contribution of this paper is the development of a mathematical framework and the analysis of optimal fractionation schedules in the presence of accelerated repopulation. We give qualitative and structural insights on the optimal fractionation scheme, with the hope that it can guide actual practice. Specifically:
\begin{enumerate}
\item We formulate a problem that includes general tumor repopulation characteristics and develop a DP approach to solve it. We choose to model accelerated repopulation implicitly by using decelerating tumor growth curves, where a larger number of tumor cells results in slower growth. Thus, faster growth is exhibited towards the end of radiation treatment, when there are fewer cells. 
\item We prove that the optimal doses are non-decreasing over time (Theorem \ref{thm3}), due to the decelerating nature of tumor growth curves. This type of result remains valid even when we allow for weekend and holiday breaks (Corollary \ref{coro1}).
\item We analyze the special structure of the problem for the case of Gompertzian tumor growth and show that it is equivalent to maximizing a discounted version of the BED in the tumor (Subsection \ref{subsubsec:Gomp}), which results in a simplified DP algorithm. 
\item We show that when there is repopulation, the optimal number of dose fractions is finite (Theorem \ref{thm4}). 
\item We find through numerical simulations that the optimal fraction sizes are approximately proportional to the instantaneous proliferation rate, suggesting larger dose fractions later in the treatment to compensate for the increased tumor proliferation. 
\end{enumerate} 

\subsection{Organization}

In Section \ref{mod}, we present the model, formulation, and DP solution approach. We also analyze the special structure of the problem for the case of Gompertzian tumor growth. In Section \ref{prop_opt}, we discuss both previously known results and the main results of this paper. Our primary conclusion is that the optimal dose fractions are non-decreasing over time. In Section \ref{num}, we present and discuss numerical results under exponential or Gompertzian growth models. In Section \ref{sec:disc}, we provide further remarks about the model under other assumptions, and discuss the results in relation to prior work. Finally, in Section \ref{sec:con}, we summarize our main findings and the most important implications. 

\section{Model, Formulation, and Solution Approach} \label{mod}

First, we discuss a model of radiation cell kill in Subsection \ref{subsec:LQ}, and a model of tumor growth in Subsection \ref{subsec:growth}. Next, we combine these two models and formulate a fractionation problem in Subsection \ref{subsec:form}. We also discuss simplifications for the case of exponential and Gompertzian tumor growth. Finally, we develop a DP framework to solve the problem in Subsection \ref{subsec:DP}.

\subsection{Model of Radiation Cell Kill} \label{subsec:LQ}

\begin{figure}
\centerline{\includegraphics[width=120mm]{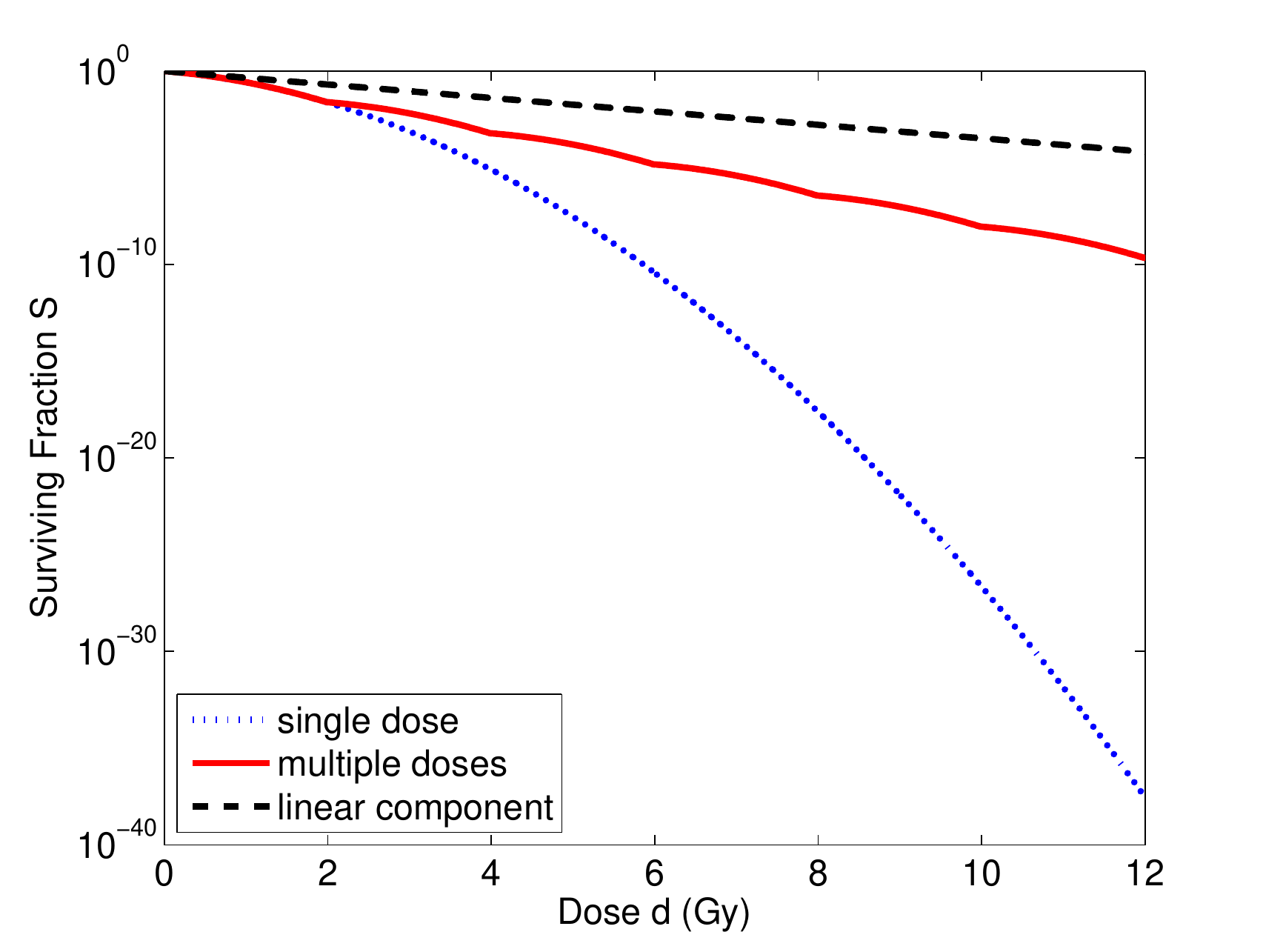}}
\caption{Illustration of the fractionation effect using the LQ model. The dotted line represents the effect of both the linear and quadratic terms resulting from applying a single total dose of radiation. The solid line corresponds to the effect of the same total dose, if it is divided into multiple individual doses, which results in a much higher survival fraction when the quadratic $\beta$ term is significant. Finally, the dashed line shows the total effect of the linear term, whether doses are applied as single or multiple individual fractions.}
\label{fig:LQ}
\end{figure}

In this subsection, we describe the radiation cell kill model without any tumor growth dynamics. We use the linear-quadratic (LQ) model (\cite{Fow1989}) to relate radiation dose and the fraction of surviving cells. This model is supported by observations from irradiating cells \emph{in vitro}. The LQ model relates the expected survival fraction $S$ (in the absence of tumor growth) after a single delivered dose $d$, in terms of two tissue parameters $\alpha$ and $\beta$, through the relation
\begin{equation*}
S = \exp(-(\alpha d + \beta d^2)).
\end{equation*}
Thus, the logarithm of the survival fraction consists of a linear component with coefficient $\alpha$ and a quadratic component $\beta$ (see the blue curve in Figure \ref{fig:LQ}). This LQ model assumes two components of cell killing by radiation; one proportional to dose and one to the square of the dose. The respective tissue-specific proportionality constants are given by $\alpha$ and $\beta$. It is possible to interpret these two components of killing as they relate to the probability of exchange aberrations in chromosomes (see \cite{HaG2006}). We illustrate this cell kill effect by plotting the logarithm of the survival fraction in Figure \ref{fig:LQ}.

For $N$ treatment days with radiation doses $d_0,d_1,\ldots,d_{N-1}$, the resulting survival fractions from each individual dose can be multiplied, assuming independence between dose effects. The resulting relation is 
\begin{equation*}
S = \exp \left(- \sum_{k=0}^{N-1} \left(\alpha d_k + \beta d_k^2 \right) \right).
\end{equation*}
The effect of the quadratic factor $\beta$, in the above equation, is that the survival fraction is larger when splitting the total dose into individual dose fractions (Figure \ref{fig:LQ}). Thus, there is an inherent trade-off between delivering large single doses to maximize cell kill in the tumor and fractionating doses to spare normal tissue. 

A common quantity that is used alternatively to quantify the effect of the radiation treatment is the \emph{biological effective dose} (BED) (\cite{Bar1982,HaG2006,OMH2009}). It is defined by
\begin{equation}
\text{BED}(d) = \frac{1}{\alpha}\left(\alpha d + \beta d^2 \right) = d\left(1+\frac{d}{\alpha/\beta}\right), \label{BED_paper9}
\end{equation}
where $\alpha/\beta$ is the ratio of the respective tissue parameters. Thus, the BED in the above definition captures the effective biological dose in the same units as physical dose. A small $\alpha/\beta$ value means that the tissue is sensitive to large doses; the BED in this case grows rapidly with increasing dose per fraction. Note that BED is related to the LQ model by setting $\text{BED}=-\ln(S)/\alpha$. In the BED model, only a single parameter, the $\alpha/\beta$ value, needs to be estimated, e.g., in \cite{MRZ2012} the $\alpha/\beta$ value is estimated for prostate cancer from radiotherapy outcomes of thousands of patients. Whenever non-standard fractionation schemes are used in a clinical setting, the BED model is typically used to quantify fractionation effects. In this paper, we frequently switch between the cell interpretation in the LQ model and the effective dose interpretation in the BED model as they both provide alternative and useful views based on context.

Based on the relation given in Equation (\ref{BED_paper9}), we define $\text{BED}_T(d)$ as the BED in a tumor when a dose $d$ is delivered, where $\abt$ is the $\alpha/\beta$ value of the tumor. We also define the total BED in the tumor from delivering doses $d_0,d_1,\ldots,d_{N-1}$ as 
\begin{equation*}
\text{BED}_T = \sum_{k=0}^{N-1} \text{BED}_T(d_k) = \sum_{k=0}^{N-1} d_k \left(1 + \frac{d_k}{[\alpha / \beta ]_T} \right). \label{bedt}
\end{equation*}
In this paper, we consider a single dose-limiting radio-sensitive organ-at-risk (OAR); this assumption is appropriate for some disease sites (e.g., for prostate cancer, the rectum could be taken as the dose-limiting organ). We assume that an OAR receives a fraction of the dose applied to the tumor. Thus, let a dose $d$ applied to the tumor result in a dose of $\gamma d$ in the OAR, where $\gamma$ is the fractional constant, also referred to as the normal tissue sparing factor, satisfying $0 < \gamma < 1$. Implicitly, this assumes a spatially homogeneous dose in the tumor and the OAR as in \cite{MTD2012}. The generalization to a more realistic inhomogeneous OAR dose distribution (\cite{UCS2013}), which leaves the main findings of this paper unaffected, is detailed in Subsection \ref{subsec:nonuniform}. The value of $\gamma$ will depend on the treatment modality and the disease site. For treatment modalities providing very conformal dose around the tumor and disease sites with the OAR not closely abutting the primary tumor, the OAR will receive less radiation and thus $\gamma$ would be a smaller. Using $\gamma d$ as the dose in the OAR and $\abo$ as the OAR $\alpha/\beta$ value, we can define the associated OAR BEDs, $\text{BED}_O(d)$ and $\text{BED}_O$, in the same way as was done for the tumor BED:
\begin{equation*}
\text{BED}_O = \sum_{k=0}^{N-1} \text{BED}_O(d_k) = \sum_{k=0}^{N-1} \gamma d_k \left(1 + \frac{\gamma d_k}{[\alpha / \beta ]_O} \right).
\end{equation*}

\subsection{Tumor Growth Model} \label{subsec:growth}

\begin{figure}
\centerline{\includegraphics[width=120mm]{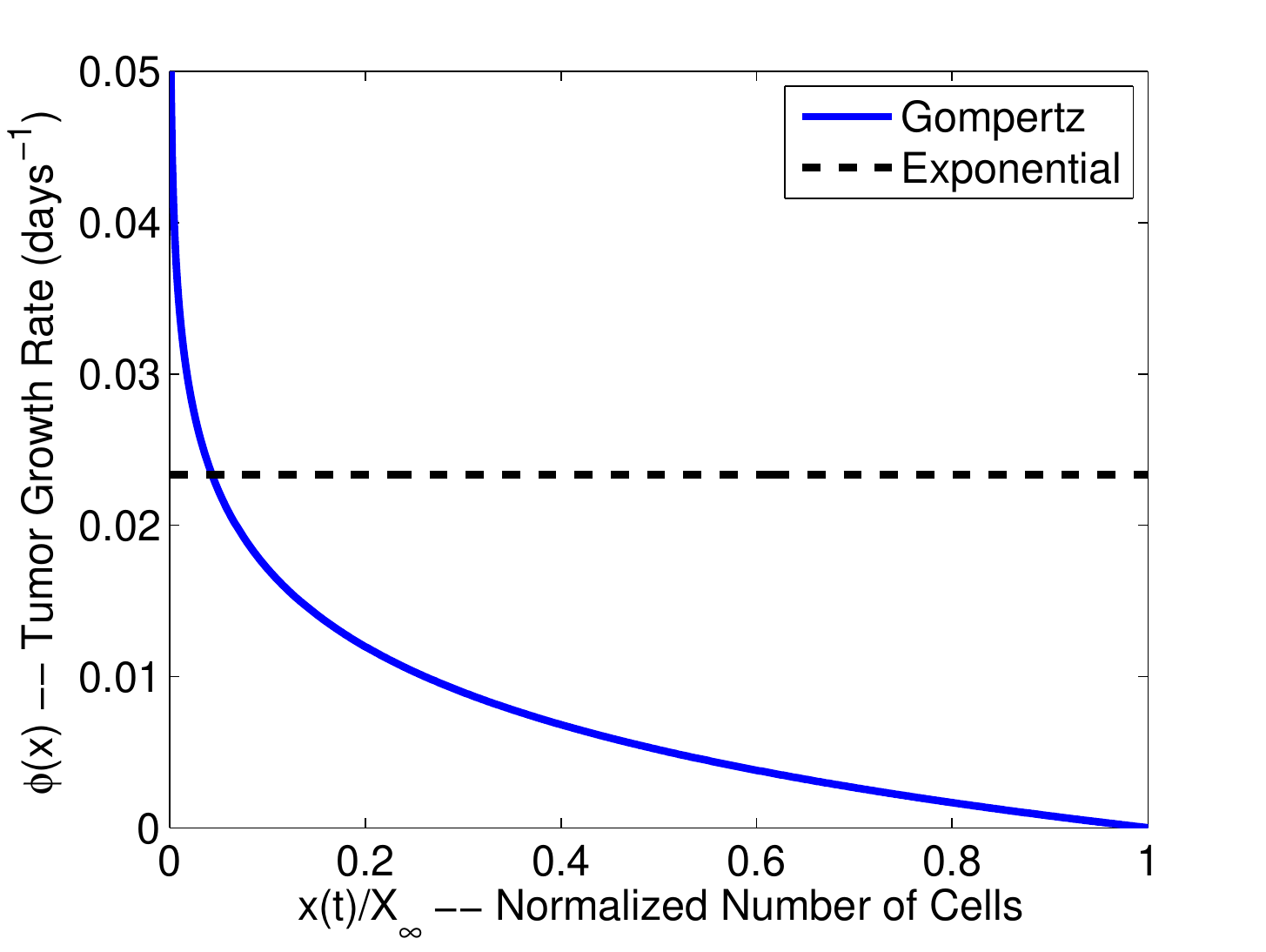}}
\caption{Tumor growth rate vs. number of tumor cells. The Gompertz equation models slower growth for larger number of cells while the exponential model assumes a constant growth rate.}
\label{fig:types_growth}
\end{figure}

In this subsection, we describe tumor growth models that will be used later to formulate a fractionation problem. We model the growth of the tumor through the ordinary differential equation (\cite{Whe1988})
\begin{equation}
\frac{1}{x(t)} \frac{\text{d}x(t)}{\text{d}t} = \phi(x(t)), \label{diffeq}
\end{equation}
with initial condition $x(0)=X_0$, where $x(t)$ is the expected number of tumor cells at time $t$. In the above equation, $\phi(x)$ represents the instantaneous tumor proliferation rate. We assume that $\phi(x)$ is non-increasing and is continuous in $x$ (for $x>0$), which implies that the solution to the above differential equation exists and is unique for any $X_0 > 0$. By choosing an appropriate functional form of $\phi$, we can describe a variety of tumor repopulation characteristics relevant for radiation therapy:

\begin{enumerate}

\item We can model \emph{exponential} tumor growth (\cite{Whe1988,YFN1993}) with a constant proliferation rate $\rho$ by choosing $\phi(x) = \rho$ . In this case, the solution $x(t)$ of the differential equation with initial condition $x(0) = X_0$ is
\begin{equation*}
x(t)=X_0 \exp (\rho t),
\end{equation*}
where $X_0$ is the initial number of cells and $\rho > 0$ is the proliferation rate. 

\item We represent \emph{accelerated} repopulation by choosing $\phi(x)$ to be a decreasing function of $x$. In this case, the instantaneous tumor proliferation rate increases when, towards the end of treatment, the number of remaining tumor cells decreases. The Gompertz model (\cite{Lai1964,NSB1976,Nor1988}) is one such decelerating tumor growth curve (Figure \ref{fig:types_growth}). For Gompertzian growth, we would simply set
\begin{equation*}
\phi(x) = b \ln \left( \frac{X_\infty}{x} \right),
\end{equation*}
where $X_0$ is the initial number of tumor cells, $X_\infty$ is the carrying capacity or the maximum number of tumor cells, and $b$ is a parameter that controls the rate of growth. The solution to the differential equation (\ref{diffeq}) with initial condition $x(0) = X_0$ is 
\begin{equation}
x(t) = X_0^{\exp (-bt)} X_\infty^{1-\exp (-bt)}, \label{gomp2}
\end{equation}
The above equation models slower repopulation for larger tumor sizes and vice versa (see Figure \ref{fig:types_growth}).  
\end{enumerate}

\subsection{Formulation} \label{subsec:form}

In this subsection, we combine the LQ model from Subsection \ref{subsec:LQ} and the tumor growth model from Subsection \ref{subsec:growth}, and formulate a fractionation problem. The aim of radiation therapy is to maximize the tumor control probability (TCP) subject to an upper limit on the normal tissue complication probability (NTCP) in the OAR (\cite{OMH2009}). A significant amount of research has been conducted to determine appropriate and better models of TCP (\cite{BrA1987,OMH2009}) and NTCP (\cite{KuB1989,Lym1985}). A common way to model normal tissue complication probability (NTCP) is as a sigmoidal function of $\text{BED}_O$ (\cite{KBB1991}). Since a sigmoidal function is monotonic in its argument, it then suffices in our model to impose an upper limit on $\text{BED}_O$. Though some studies have raised concerns (\cite{TTT1990}), TCP has been widely modeled using Poisson statistics (\cite{MuG1961,Por1980a,Por1980b}), under which
\begin{equation*}
\text{TCP} = \exp(-X_{N-1}^+),
\end{equation*}
where $X_{N-1}^+$ is the expected number of tumor cells surviving after the last dose of radiation. In this case, maximizing the TCP is equivalent to minimizing $X_{N-1}^+$. We now define
\begin{equation*}
Y_{N-1}^+ = \ln (X_{N-1}^+)/\alpha_T,
\end{equation*}
where $\alpha_T$ is a tumor tissue parameter associated with the linear component of the LQ model. Note that the definition of $Y$ is analogous to the definition of the BED. It has units of radiation dose; thus differences in $Y$ can be interpreted as differences in effective BED delivered to the tumor. We choose to primarily work with this logarithmic version because of this interpretation.

For the rest of the paper, we focus on the equivalent problem of minimizing $Y_{N-1}^+$ subject to an upper limit on $\text{BED}_O$. The problem is stated mathematically as
\begin{equation}
\underset{\{ d_i \geq 0 \} }{\text{minimize}} \quad Y_{N-1}^+ \quad \text{s.t.} \quad \text{BED}_O \leq c, \label{opt_prob2}
\end{equation}
where $c$ is a prespecified constant. There is no guarantee of the convexity of the objective, and thus, this problem is non-convex. Note that the feasible region is non-empty because a possible feasible schedule is a dose of zero for all treatment sessions. We claim that the objective function in (\ref{opt_prob2}) attains its optimal value on the feasible region. As a function of the dose fractions $d_k$, it can be seen that $Y_{N-1}^+$ is continuous. Furthermore, the constraint on $\text{BED}_O$ ensures the feasible region is compact. Thus, the extreme value theorem ensures that the objective attains its optimal value on the feasible region.

\begin{figure}
\centerline{\includegraphics[width=120mm]{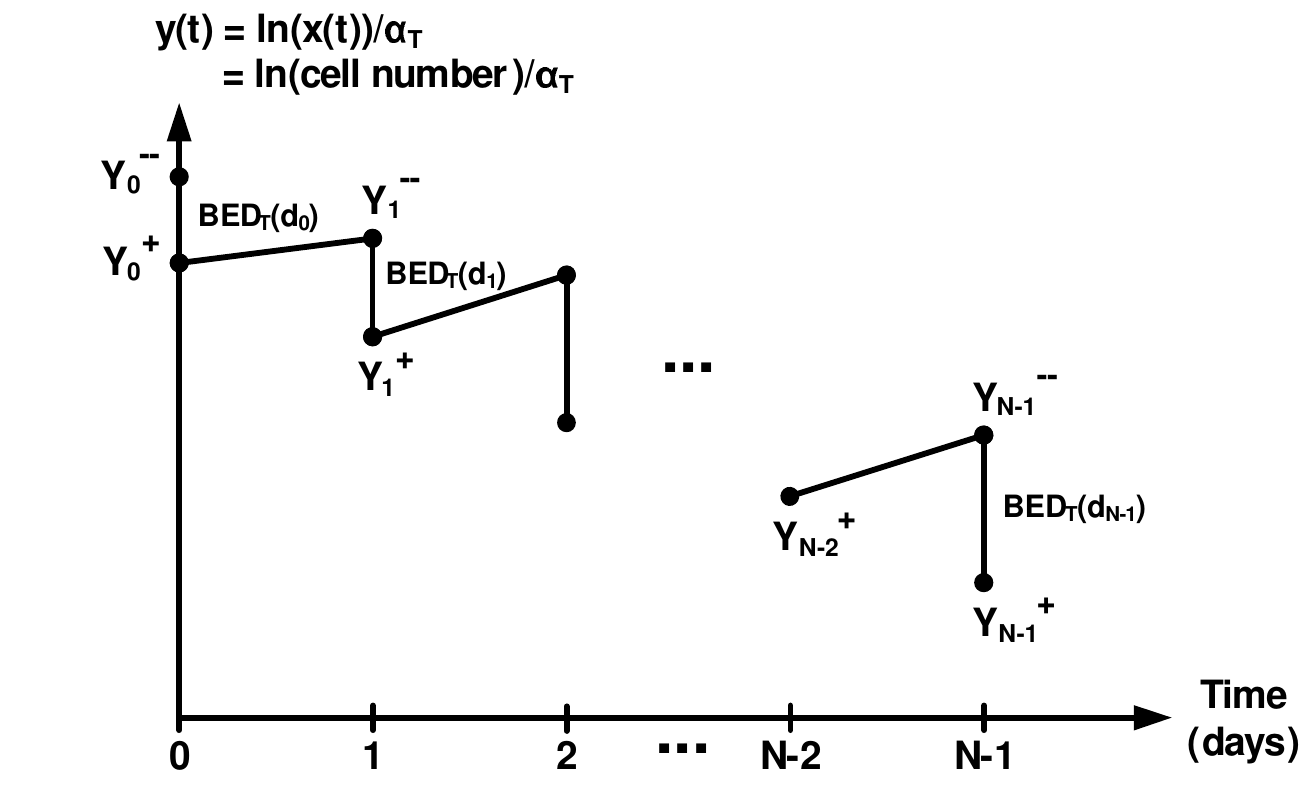}}
\caption{Schematic illustration of the expected number of tumor cells over the course of treatment. The effect of radiation dose $d$ is a reduction, proportional to $\text{BED}_T(d)$, in the log of the number of cells. }
\label{fig:cells}
\end{figure}

We now describe the dynamics of the expected number of tumor cells during the course of treatment (Figure \ref{fig:cells}). We assume that a sequence of $N$ doses $d_0, d_1, \ldots, d_{N-1}$ is delivered at integer times, i.e., time is measured in days. The survival fraction of cells from delivering these radiation doses is described by the LQ model in Subsection \ref{subsec:LQ}. If $X_i^-$ and $X_i^+$ are the numbers of tumor cells immediately before and after delivering the dose $d_i$, we will have $X_i^+ = X_i^- \exp(-(\at d_i + \bt d_i^2))$. For the logarithmic versions $Y_i^+$ and $Y_i^-$, we have for integer times 
\begin{equation*}
Y_i^+ = Y_i^- - \text{BED}_T(d_i). \label{lq2}
\end{equation*}
For non-integer times in $[0,N-1]$, the tumor grows according to the differential equation (\ref{diffeq}) with proliferation rate $\phi(x)$, as described in Subsection \ref{subsec:growth}. For convenience, we denote by $F(\cdot)$ the resulting function that maps $Y^-$ to $Y^+$ when using the growth differential equation (\ref{diffeq}). Thus, we have 
\begin{equation*}
Y_{i+1}^- = F(Y_i^+),
\end{equation*}
for $i=0,1,\ldots,N-2$.

\subsubsection{Exponential Tumor Growth with Constant Proliferation Rate}

For the exponential tumor growth model where $\phi(x)=\rho$, the rate of repopulation $\rho$ does not change with tumor size (see Figure \ref{fig:types_growth}). Assuming $\rho$ represents a measure of growth per unit day (or fraction), the number of tumor cells is multiplied by a factor of $\exp (\rho)$ after every fraction. Or, equivalently, a constant factor is added to the logarithmic cell number due to tumor growth in between treatment days. Since there are $N$ dose fractions and $N-1$ days of repopulation in between, it can be seen that the optimization problem (\ref{opt_prob2}) simplifies to
\begin{equation}
\underset{\{ d_i \geq 0 \} }{\text{minimize}} \quad Y_0 + \frac{1}{\at}(N-1) \rho - \text{BED}_T \quad \text{s.t.} \quad \text{BED}_O \leq c. \label{opt_prob6}
\end{equation}
Here, the effect of tumor repopulation is captured in the term $(N-1) \rho / \at$.

\subsubsection{Gompertzian Tumor Growth} \label{subsubsec:Gomp}

For the case of Gompertzian tumor growth where $\phi(x) = b \ln \left(X_\infty / x \right)$, we can also significantly simplify the formulation. As will be seen in the next subsection, this results in a simplified DP approach to determine an optimal fractionation schedule. We will now derive an explicit expression for the expected number of tumor cells at the end of treatment. We claim that 
\begin{equation}
Y_i^+ = \frac{1}{\at} \ln \left( x(i) \right) - \sum_{k=0}^i \exp [-b (i-k)] \text{BED}_T(d_k), \label{gomp_rec}
\end{equation} 
for $i = 0, 1, \ldots,N-1$, where $x(t)$ represents the expected number of tumor cells in the absence of radiation treatment, and which is given by the expression in Equation (\ref{gomp2}). (In particular, $x(t)$ should not be confused with the number of tumor cells when the tumor is treated.) Equation (\ref{gomp_rec}) holds for $i=0$ because $x(0) = X_0$ and $Y_0^+ = \ln(X_0)/\at - \text{BED}_T(d_0)$. For the inductive step, suppose that Equation (\ref{gomp_rec}) holds true. From the Gompertzian growth equation (\ref{gomp2}), and assuming that the time interval between fractions is one day ($t=1$), we can write the function $F$ that maps $Y_i^+$ to $Y_{i+1}^-$ as
\begin{equation*}
Y_{i+1}^- = F(Y_i^+) = \exp (-b) Y_i^+ + (1-\exp (-b)) \frac{\ln (X_\infty)}{\at}.
\end{equation*}
Incorporating the growth and the radiation dose from $d_{i+1}$, we find
\begin{align*}
Y_{i+1}^+ &= F(Y_i^+) - \text{BED}_T(d_{i+1}) \\
&= \frac{1}{\at} \ln \left( x(i)^{\exp (-b)} X_\infty^{(1-\exp (-b))} \right) - \sum_{k=0}^{i+1} \exp [-b (i+1-k)] \text{BED}_T(d_k) \\
&= \frac{1}{\at} \ln \left( x(i+1) \right) - \sum_{k=0}^{i+1} \exp [-b (i+1-k)] \text{BED}_T(d_k),
\end{align*}
completing the inductive step. This results in the optimization problem 
\begin{equation*}
\begin{aligned}
& \underset{\{ d_i \geq 0 \}}{\text{minimize}}
& & \frac{1}{\at} \ln \left( x(N-1) \right) - \sum_{k=0}^{N-1} \exp [-b (N-1-k)] \text{BED}_T(d_k) \\  
& \; \; \text{s. t.}
& & \text{BED}_O \leq c, \label{opt_prob4}
\end{aligned}
\end{equation*}
The interesting aspect of the above optimization problem is that it is essentially a maximization of a discounted sum of the terms $\text{BED}_T(d_k)$. Because the weighting term gives larger weight to later fractions, we can conjecture that the optimal fractionation scheme will result in larger fraction sizes towards the end of treatment. This is in contrast to the exponential tumor growth model for which there is no accelerated repopulation, and the $\text{BED}_T(d_k)$ terms are weighted equally.

\subsection{Dynamic Programming Approach} \label{subsec:DP}

In order to get from the initial $Y_0$ to $Y_{N-1}^+$, one recursively alternates between applying a dose $d$ and the growth function $F$. That is, $Y_{N-1}^+$ takes the form
\begin{equation}
Y_{N-1}^+ = F( \cdots F( F(Y_0 - \text{BED}_T(d_0)) - \text{BED}(d_1) ) \cdots ) - \text{BED}(d_{N-1}). \label{obj}
\end{equation}
Such a recursive formulation lends itself naturally to a Dynamic Programming (DP) approach. We can solve the optimization problem by recursively computing an optimal dose backwards in time. Note that although non-linear programming methods can also be used, there is no guarantee of the convexity of (\ref{obj}), and thus, such methods might only provide a local optimum. A global optimum, on the other hand, is guaranteed if a DP approach is used. 

In order to determine the dose $d_k$, we take into account $Y_{k-1}^+$ and the cumulative BED in the OAR from delivering the prior doses, which we define as 
\begin{equation*}
z_k = \sum_{i=0}^{k-1} \text{BED}_O(d_i).
\end{equation*}
Here, $Y_{k-1}^+$ and $z_k$ together represent the state of the system because they are the only relevant pieces of information needed to determine the dose $d_k$. We do not include a cost per stage; instead, we include $Y_{N-1}^+$ in a terminal condition. In order to ensure that the constraint on $\text{BED}_O$ is satisfied, we also assign an infinite penalty when the constraint is violated. The Bellman recursion to solve the problem is:
\begin{equation*}
	J_{N}(Y_{N-1}^+,z_N) = \left\{
	\begin{array}{lll}
		Y_{N-1}^+,  					&	&\text{if $z_N \leq c$},\\
		\infty,			&	&\text{otherwise},
	\end{array} \right.
	\label{J_last}
\end{equation*}
\begin{equation}
J_{k}(Y_{k-1}^+,z_k) = \underset{d_k \geq 0}{\text{min}} \left[ J_{k+1}(F(Y_{k-1}^+) - \text{BED}_T(d_k), z_k + \text{BED}_O(d_k)) \right], \label{J_k}
\end{equation}
for $k=N-1,N-2,\ldots,1$. The initial equation for time $0$, given below, is slightly different because there is no prior tumor growth (see Figure \ref{fig:cells}):
\begin{equation*}
J_{0}(Y_0^-,z_0) = \underset{d_0 \geq 0}{\text{min}} \left[ J_{1}(Y_0^- - \text{BED}_T(d_0), z_0 + \text{BED}_O(d_0)) \right]. \label{J_0}
\end{equation*}

For the exponential and Gompertzian growth cases, the DP approach simplifies and only requires the single state $z_k$. Below, we discuss the approach for the Gompertzian case only; for the exponential case, as will be discussed in Subsection \ref{subsec:exist}, an optimal fractionation scheme can be characterized in closed form. For simplicity, we use additive costs per stage this time. The simplified algorithm is 
\begin{equation*}
	J_{N}(z_{N}) = \left\{
	\begin{array}{lll}
		\frac{1}{\at} \ln \left( x(N-1) \right),  					&	&\text{if $z_{N} \leq c$},\\
		\infty,			&	&\text{otherwise},
	\end{array} \right.
	\label{J_last2}
\end{equation*}
\begin{equation}
J_{k}(z_k) = \underset{d_k \geq 0}{\text{min}} \Big[ - \exp [-b (N-1-k)] \text{BED}_T(d_k) + J_{k+1}(z_k + \text{BED}_O(d_k)) \Big], \label{J_k2}
\end{equation}
for $k=N-1,N-2,\ldots,0$. For numerical implementation, the state variables need to be discretized and the tabulated values stored. For evaluating the cost-to-go function $J_k$ at any non-discretized values, an interpolation of appropriate discretized values can be used for increased accuracy. 

\section{Properties of an Optimal Fractionation Schedule} \label{prop_opt}

In Subsection \ref{subsec:exist}, we discuss previously known results. These primarily concern the characterization of optimal fractionation schedules in the absence of accelerated repopulation. The purpose of this subsection is to provide the essential results that are scattered in the literature in different papers, and formalize them in the framework of this paper. In Subsection \ref{subsec:new}, we summarize our main results. The proofs of the results are provided in the Appendix. 

\subsection{Previously Known Results} \label{subsec:exist}

The set of all optimal solutions to the fractionation problem in the absence of repopulation is characterized in the following theorem, published in \cite{MTD2012}. 

\begin{theorem}
Let $N$ be fixed. In the absence of repopulation (i.e., $\phi(x)=0$), an optimal fractionation schedule can be characterized in closed form. If $\abo \geq \g \abt$, an optimal solution is to deliver a single dose equal to
\begin{equation}
d_j^* = \frac{\abo}{2\g} \left[ \sqrt{1 + \frac{4 c}{\abo}} - 1 \right] \label{mizuta1}
\end{equation}
at an arbitrary time $j$ and deliver $d_i = 0$ for all $i \neq j$. This corresponds to a hypo-fractionation regimen. If $\abo < \gamma \abt$, the unique optimal solution consists of uniform doses given by
\begin{equation}
d_j^* = \frac{\abo}{2\g} \left[ \sqrt{1 + \frac{4 c}{N \abo}} - 1 \right], \label{mizuta2}
\end{equation}
for $j = 0, 1, \ldots, N-1$. This corresponds to a hyper-fractionation regimen, i.e., a fractionation schedule that uses as many treatment days as possible. 
\label{thm1}
\end{theorem}

The above theorem states that if $\abo \geq \g \abt$, a single radiation dose is optimal, i.e., the optimal number of fractions $N^*$ is $1$. On the other hand, if $\abo$ is small enough, i.e., the OAR is sensitive to large doses per fraction, so that $\abo < \gamma \abt$, it is optimal to deliver the same dose during the $N$ days of treatment. Because taking larger $N$ only results in extra degrees of freedom, $N^* \rightarrow \infty$ in this case. However, this is clearly not realistic and is an artifact of modeling assumptions. We will show in the next subsection that including tumor repopulation results in a finite $N^*$. In the following remark, we comment on the result and the model assumptions. 

\begin{remark} \label{rem:single}
Our ultimate goal is to understand the effect of accelerated repopulation over the duration of treatment. Thus, we are primarily interested in the hyper-fractionation case, with $\abo < \g \abt$, which will hold for most disease sites. The case where $\abo \geq \g \abt$ needs careful consideration since the validity of the model may be limited if $N$ is small and the dose per fraction is large. The condition $\abo \geq \g \abt$ would be satisfied for the case of an early responding OAR tissue (see Subsection \ref{subsec:tissue} for further details); however, we have not included repopulation and repair effects into the BED model, which could play an important role for such tissue. Another aspect that requires consideration is whether doses should be fractionated to permit tumor re-oxygenation between treatments. Thus, without extensions to our current model, it may be better to set a minimum number of fractions (e.g., 5) in the case where $\abo \geq \g \abt$.
\end{remark}In the next remark, we discuss how the above theorem can be generalized in the case of exponential tumor growth. 

\begin{remark}
For the problem including exponential tumor growth with $\phi(x) = \rho$, there is only an additive term $(N-1) \rho / \at$ in the objective (\ref{opt_prob6}), which is independent of the dose fractions. Thus, for a fixed $N$, the result from Theorem \ref{thm1} still holds for the exponential growth case. 
\end{remark}

It turns out that one can also characterize the optimal number of fractions in closed form for the exponential tumor growth case. The result is consistent and similar to the work in a few papers (\cite{WKO1977,JTD1995,ADJ2004}) though it is interpreted differently here. Our statement below is a slight generalization in that we do not assume uniform dose per day a priori.  

\begin{theorem}
The optimal number of fractions $N^*$ for exponential growth with constant proliferation rate $\rho$ (i.e., $\phi(x)=\rho$) is obtained by the following procedure:
\begin{enumerate}
\item If $\abo \geq \g \abt$, then $N^*=1$.
\item If $\abo < \g \abt$, then
\begin{enumerate}
\item Compute $N_c = A \left( \sqrt{\frac{\left( \rho + B \right)^2}{\rho \left( \rho + 2B \right)}} - 1\right)$, where $A = \frac{2 c}{\abo}$, $B = \frac{\alpha_T \abo}{2\g} \left( 1 - \frac{\abo}{\g \abt} \right)$.
\item If $N_c<1$, then $N^*=1$. Otherwise, evaluate the objective at $\lfloor N_c \rfloor$ and $\lceil N_c \rceil$, where $\lfloor \cdot \rfloor$ and $\lceil \cdot \rceil$ are the floor and ceiling operators, respectively, and let the optimum $N^*$ be the one that results in a better objective value.
\end{enumerate}
\end{enumerate} 
\label{thm2}
\end{theorem}

This result also makes sense in the limiting cases. When approaching the case of no repopulation, i.e., $\rho \rightarrow 0$, we see that $N_c \rightarrow \infty$, and the optimum $N^*$ approaches infinity. When $\rho \rightarrow \infty$, we see that $N_c \rightarrow 0$, meaning that the optimum $N^*$ is a single dose. Recall that if $\abo < \g \abt$, $N^* \rightarrow \infty$ in the absence of repopulation. For the case of exponential repopulation with constant but positive rate, the above result shows that the optimum $N$ is finite. Indeed, even for general tumor growth characteristics, we will show in the next subsection that as long as there is some repopulation, the optimal number of fractions will be finite.  

\subsection{Main Results of This Paper} \label{subsec:new}

We begin with two lemmas. The first one simply states that the constraint on $\text{BED}_O$ is binding; intuitively, this is because of the assumption that $\phi(x)>0$ (implying the growth function $F(\cdot)$ is increasing) and the fact that the BED function is monotone in dose.
\begin{lemma}
Assume that $\phi(x)>0$ for all $x>0$. Then, the constraint on $\text{BED}_O$ in (\ref{opt_prob2}) will be satisfied with equality.
\label{lem3}
\end{lemma}

\begin{lemma}
Assume that $\phi(x)>0$ for all $x>0$ and that $\phi(x)$ is a non-increasing function of $x$. Suppose that $i < j$ and that we apply the same sequence of doses $d_{i+1},\ldots,d_j$, starting with either $\yip$ or $\widetilde{Y}_{i}^+$. If $\yip<\widetilde{Y}_{i}^+$, then $\yjp<\widetilde{Y}_{j}^+$ and $\widetilde{Y}_{j}^+ - \yjp \leq \widetilde{Y}_{i}^+ - \yip$.
\label{lem2}
\end{lemma}

\noindent Assuming that the same sequence of doses are applied, Lemma \ref{lem2} above states a monotonicity and a contraction type property when mapping the expected number of cells from one point in time to another. Next, we state the main theorem. 

\begin{theorem}
Let us fix the number of treatment days $N$. Assume that there is always some amount of repopulation, i.e., $\phi(x)>0$ for all $x>0$, and that the instantaneous tumor growth rate $\phi(x)$ is non-increasing as a function of the number of cells $x$. If $\abo \geq \g \abt$, then it is optimal to deliver a single dose equal to (\ref{mizuta1}) on the last day of treatment. This corresponds to a hypo-fractionation regimen. If $\abo < \g \abt$, then any optimal sequence of doses is non-decreasing over the course of treatment. That is, these doses will satisfy $d_0^* \leq d_1^* \leq \cdots \leq d_{N-1}^*$.
\label{thm3}
\end{theorem}

\noindent For the case where $\abo \geq \g \abt$, we take note of Remark \ref{rem:single} again. It is reasonable that an optimal solution uses the most aggressive treatment of a single dose as this is the case even without repopulation. The next remark gives an intuitive explanation for why in the fixed $N$ case it is optimal to deliver only on the last treatment day, and shows the optimal number of fractions $N^*$ is 1 when $\abo \geq \g \abt$. 

\begin{remark}
A single dose delivered on the last treatment day is optimal when $\abo \geq \g \abt$ because it is better to let the tumor grow slowly during the course of treatment rather than stimulate accelerated growth by treating it earlier. However, this does not mean it is optimal to wait forever before treating the patient. Starting with a given initial number of tumor cells, it is clear that treating a single dose with a smaller $N$ results in a better cost. This is because the tumor grows for a shorter duration of time. Thus, it follows that when $\abo \geq \g \abt$, the optimal number of fractions is $N^* = 1$.
\end{remark}

In the interesting case when $\abo < \g \abt$, the doses must increase over time. Intuitively, due to the decreasing property of $\phi(x)$ as a function of $x$, the tumor grows at a faster rate when its size becomes smaller over the course of treatment; higher doses are then required to counter the increased proliferation. An interchange argument is used to prove the above theorem.

The next theorem states that as long as the repopulation rate cannot decrease to zero, the optimal number of fractions $N^*$ is finite.  

\begin{theorem}
Suppose that there exists $r>0$ such that $\phi(x)>r$ for all $x>0$. Then, there exists a finite optimal number of fractions, $N^*$. 
\label{thm4}
\end{theorem}

Typically, in a clinical setting, dose fractions are not delivered during weekend and holiday breaks. The following remark explains how such breaks can be included in our formulation. 
\begin{remark} \label{rem_weekend}
We can adjust the fractionation problem by setting $N$ to be the total number of days, including weekend and holiday breaks. For days in which a treatment is not administered, the dose fraction is set to $0$. And, for all other days, the DP algorithm (\ref{J_k}) is used as before to determine the optimal fractionation schedule. 
\end{remark}

The following corollary shows that the structure of an optimal solution is still similar to that described in Theorem \ref{thm3} even when including holidays into the formulation and/or fixing some dose fractions. 

\begin{corollary}
Including breaks and/or fixing the dose in some fractions does not change the structure of the optimized dose fractions as described in Theorem \ref{thm3}. Fix $N$. Assume that $\phi(x)>0$ for all $x>0$ and that $\phi(x)$ is non-increasing in $x$. If $\abo \geq \g \abt$, then an optimal solution is to deliver a single dose on the last deliverable day that is not a break and not a fixed dose fraction. If $\abo < \g \abt$, then an optimal sequence of doses, excluding breaks and fixed dose fractions, is non-decreasing over the course of treatment. \label{coro1}
\end{corollary}

Thus, according to the above corollary, if $\abo < \g \abt$ and we introduce weekend breaks, the dose on Monday will in general be larger than the dose on the Friday of the previous week. In order to understand the intuitive reason why this is so, consider what would happen if the Friday dose were in fact larger. The tumor would then be growing at a faster rate over the weekend; but then it would make better sense to deliver a higher Monday dose instead. This result is different from the numerical results in \cite{WCW2000}. Of course, the model was different in that paper because the primary motivation was to determine the effect of the varying tumor sensitivity over the course of treatment, not to counter accelerated repopulation. 

When we assume an exponential growth model, we can fully characterize the optimal solution, which is done in the following remark. 

\begin{remark}
For the exponential growth model where $\phi(x) = \rho$, if $\abo < \g \abt$, we claim that the optimal solution would be to deliver uniform doses at each treatment session, that is, on days that are not breaks and that do not have fixed dose fractions. Let us adjust the number of days $N$ appropriately, and set the dose fractions to $0$ for the breaks. Then, for a fixed $N$, as seen from the objective in (\ref{opt_prob6}), the only relevant term that determines the dose fractions is $-\text{BED}_T$. Thus, for a fixed $N$, the arguments from Theorem \ref{thm1} hold and uniform doses are optimal. Now, we discuss the effect of breaks on the optimal number of fractions. If the number of breaks is simply a fixed number, the optimal number of days $N^*$, including holiday breaks, would be still given by the expression in Theorem \ref{thm2}. This is because the derivative of the objective given in (\ref{opt_prob6}) as a function of $N$ would remain unchanged. If the number of breaks is non-decreasing in the number of treatment days (as would be the case with weekend breaks), the expression for $N^*$ given in Theorem \ref{thm2} would not necessarily hold. Deriving a closed form formula for $N^*$ for this case appears to be tedious, if not impossible. We suggest therefore to exhaustively evaluate the objective given in (\ref{opt_prob6}) for reasonable values of $N$ to obtain a very good, if not optimal, choice of the number of treatment days. For a given break pattern, we could also optimize the starting day of treatment by appropriately evaluating the objective in (\ref{opt_prob6}).
\end{remark}

\section{Numerical Experiments}\label{num}

In Subsection \ref{subsec:exp}, we calculate the optimal treatment duration while assuming exponential tumor growth with varying rates of repopulation. In Subsections \ref{subsec:accel} and \ref{subsec:accel_smallab}, we model accelerated repopulation using Gompertzian growth and numerically calculate the resulting optimal fractionation scheme. Finally, in Subsection \ref{subsec:weekend}, we evaluate the effect of weekend breaks. For all of our numerical simulations, we used MATLAB on a 64-bit Windows machine with 4GB RAM. 

\subsection{Faster tumor growth suggests shorter overall treatment duration} \label{subsec:exp}

\begin{figure}
\centerline{\includegraphics[width=150mm]{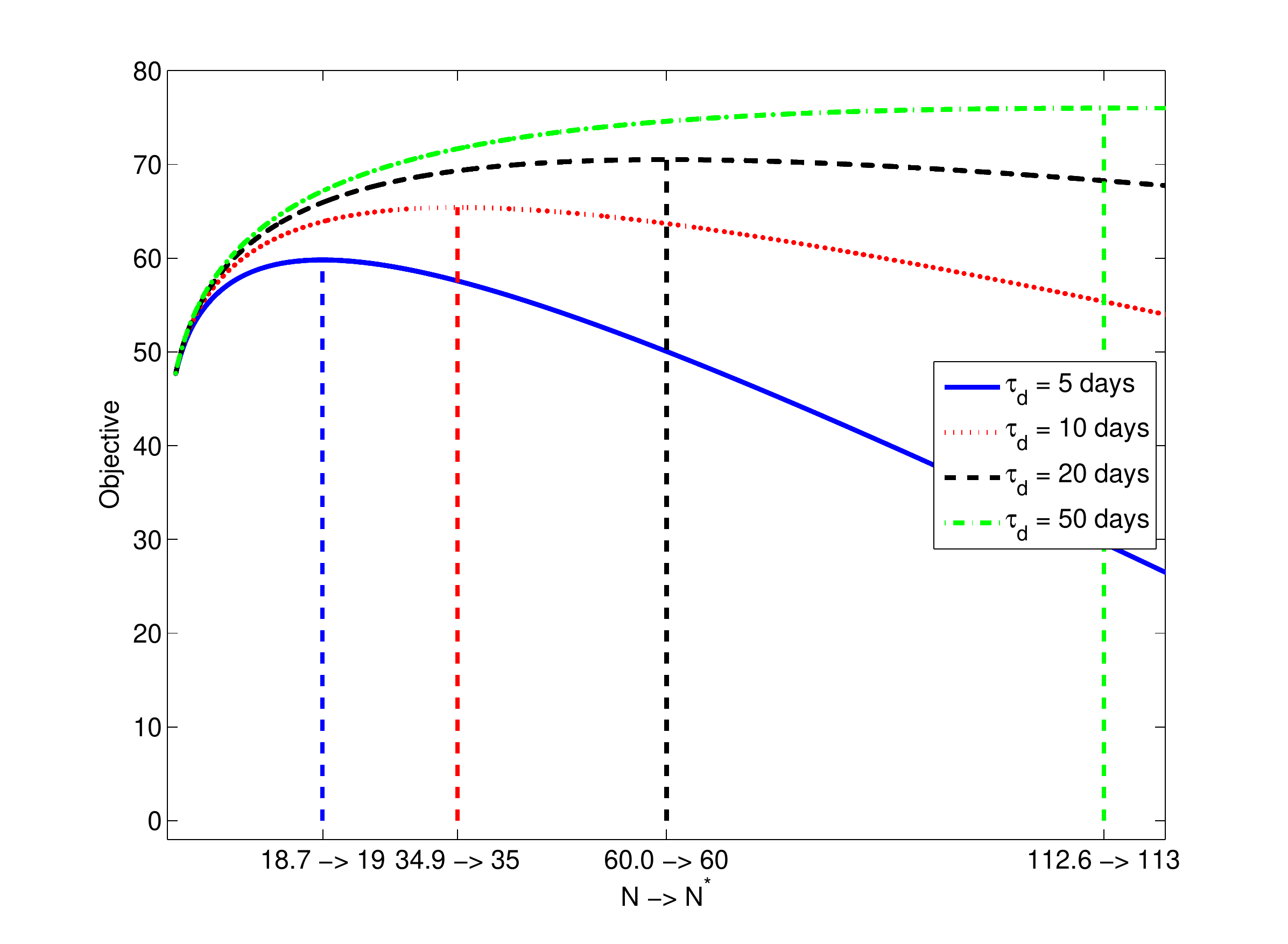}}
\caption{Dependence of the optimal value of the objective function for different choices of the number of fractions, assuming exponential tumor growth. The optimal number of treatment days is fewer for faster growing tumors. The expression in Theorem \ref{thm2} was used to generate this plot. The objective was evaluated at the floor and ceiling of the continuous optimum $N$ to obtain the actual optimum $N^*$.}
\label{fig:plotN_new}
\end{figure}

We use realistic choices of radiobiological parameters in order to assess the effect of various rates of tumor growth on the optimal number of treatment days. Here, we assume exponential growth with a constant rate of repopulation. We use $\abt = 10$ Gy, $\abo = 3$ Gy, and $\at = 0.3$ Gy$^{-1}$ for the tissue parameters; these are appropriate standard values (\cite{GuL2003,HaG2006}). We consider a standard fractionated treatment as reference, i.e., a dose of $60$ Gy delivered to the tumor in 30 fractions of $2$ Gy. For the above choice of $\alpha$/$\beta$ values and $\g = 0.7$, this corresponds to an OAR BED of 61.6 Gy, which we use as the normal tissue BED constraint $c$. In order to choose appropriate values for the proliferation rate $\rho$, we relate it to the tumor doubling time $\tau_d$. Since $\tau_d$ represents the time it takes for the tumor to double in size, we set $\exp(\rho t) = 2^{t/\tau_d}$, resulting in the following relation:
\begin{equation*}
\rho = \frac{\ln(2)}{\tau_d}.
\end{equation*}
For human tumors, the doubling time can range from days to months (\cite{WTM1988,KiT2005}), depending on the particular disease site. We observe that for the parameters assumed above, the optimal number of treatment days is smaller for faster growing tumors (Figure \ref{fig:plotN_new}). 

The objective value plotted in Figure \ref{fig:plotN_new} is 
\begin{equation*}
\text{BED}_T - \frac{1}{\at} (N-1) \rho.
\end{equation*}
For the reference treatment ($N=30$) and $\alpha_T=0.3$, the decrease in the tumor BED due to the second term $(N-1) \rho / \at$ evaluates to about $1.3$ Gy for a slowly proliferating tumor with doubling time $\tau_d=50$. This is small compared to $\text{BED}_T=72$. For a fast proliferating tumor with doubling time $\tau_d=5$, the correction $(N-1) \rho / \at$ is about $13.4$ Gy and becomes more important. Thus, smaller values of $N$ are suggested for faster proliferating tumors.

\subsection{Accelerated repopulation suggests increasing doses towards the end of treatment} \label{subsec:accel}

\begin{figure}
\centerline{\includegraphics[width=120mm]{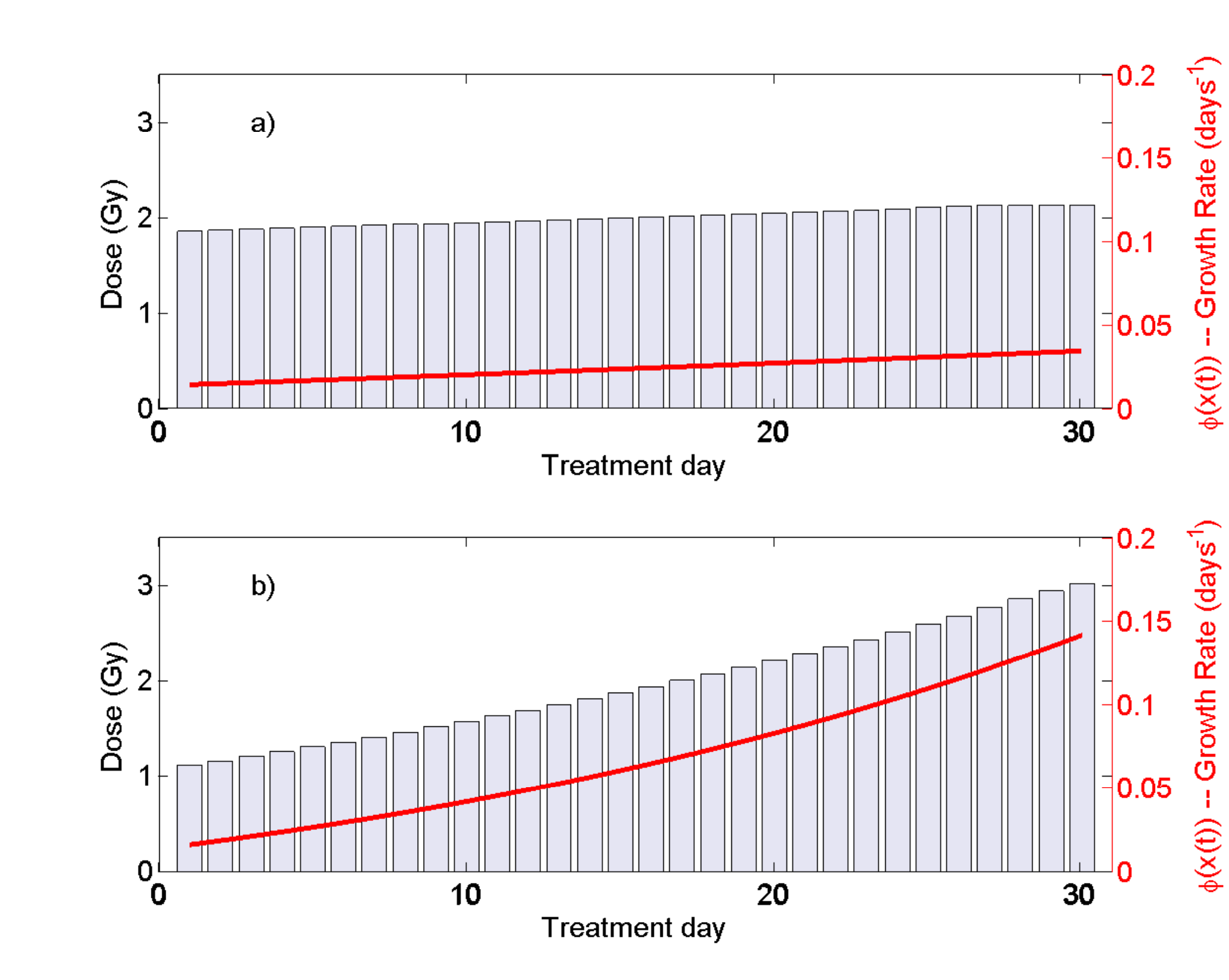}}
\caption{Optimal fractionation for accelerated repopulation. Plot a) shows the optimal fractionation schedule for a slowly proliferating tumor and plot b) for a fast one. The doubling time for the reference treatment begins at $\tau_d = 50$ days and decreases approximately to a) $20$ and b) $5$ days, respectively, at the end of treatment. The plotted $\phi(x)$ is the resulting instantaneous proliferation rate after the delivery of each dose fraction.}
\label{fig:plot1}
\end{figure}

One way to model accelerated repopulation is to use decreasing tumor growth curves (see Figure \ref{fig:types_growth}). We model this behavior by using the Gompertzian tumor growth model and solve the fractionation problem by using the simplified DP equation (\ref{J_k2}). For numerical implementation of the DP algorithm, we discretize the state into 500 points for each time period. When evaluating the cost-to-go function for values in between discretization points, we use linear interpolation. We illustrate optimal fractionation schemes for both slow and fast proliferating tumors. For a slowly proliferating tumor, we choose the parameters $X_0 = 4 \times 10^6$, $X_\infty = 5 \times 10^{12}$, and $b = \exp(-6.92)$ by fixing $X_\infty$ (to be on the order of the value given in \cite{Nor1988} for breast cancer) and manually varying $X_0$ and $b$ so that the doubling time for the reference treatment starts at $50$ days in the beginning and decreases to $20$ days at the end of treatment. For a fast proliferating tumor, we adjust the parameters accordingly so that the doubling time goes from $50$ days to $5$ days: $X_0 = 6 \times 10^{11}$, $X_\infty = 5 \times 10^{12}$, and $b = \exp(-5.03)$. Such fast growth is not atypical; from clinical data on squamous cell carcinomas of the head and neck, it is possible for the doubling time to decrease from 60 days for an unperturbed tumor to 4 days, though there could be an initial lag period of constant repopulation (\cite{WTM1988}). As before, we use the parameters $\abt = 10$ Gy, $\abo = 3$ Gy, $\at = 0.3$ Gy$^{-1}$, $c=61.6$ Gy, and $\g = 0.7$. As shown in Figure \ref{fig:plot1}, for a fast proliferating tumor, the sequence of radiation doses increase from $1$ Gy to $3$ Gy, which is a significant difference from the standard treatment of $2$ Gy per day for $30$ days. For a slow proliferating tumor, the doses closely resemble standard treatment and only increase slightly over the course of treatment. Note that the optimal fractionation scheme distributes the doses so that they are approximately proportional to the instantaneous proliferation rate, $\phi(x)$. The plotted $\phi(x)$ in Figure \ref{fig:plot1} is the resulting instantaneous proliferation rate after the delivery of each dose fraction. For the reference treatment of $2$ Gy per day with $N=30$, in the case of a fast proliferating tumor, the objective $Y_{N-1}^+$ is $26.03$. The objective $Y_{N-1}^+$ for the optimal fractionation scheme in plot b) of Figure \ref{fig:plot1} is $25.41$. This is a change of about $2.4 \%$ in $Y_{N-1}^+$ and $17.0 \%$ change in $X_{N-1}^+$ in comparison to the reference treatment. It is not straightforward to make a meaningful statement about the improvement in tumor control simply based on these values. However, we can say that even a small improvement in tumor control for a specific disease site can make a significant impact because of the large number of patients treated with radiation therapy every year. 

In principle, running the DP algorithm for every possible value of $N$ would give us the optimal number of fractions. We choose to run the algorithm for $N=1,2,\ldots,100$. For each run of the DP algorithm for a fixed $N$, it takes on average about $7$ seconds using MATLAB on a 64-bit Windows machine with 4GB RAM and a Intel i7 2.90GHz chip. We find $N=79$ results in the best cost for the slowly proliferating tumor and $N=38$ for the fast proliferating tumor. For the slowly proliferating tumor, a very small fraction of the tumor's cells remain regardless of whether $N=79$ or $N=30$, and for the fast proliferating tumor, setting $N=30$ instead of $N=38$ results only in a change of $0.7 \%$ in $Y_{N-1}^+$. Thus, for practical purposes, it is reasonable to set $N=30$ because more dose fractions could mean, among other factors, patient inconvenience and further cost.

\subsection{Smaller $\alpha/\beta$ value of the tumor results in larger changes in the fractionation schedule} \label{subsec:accel_smallab}

\begin{figure}
\centerline{\includegraphics[width=130mm]{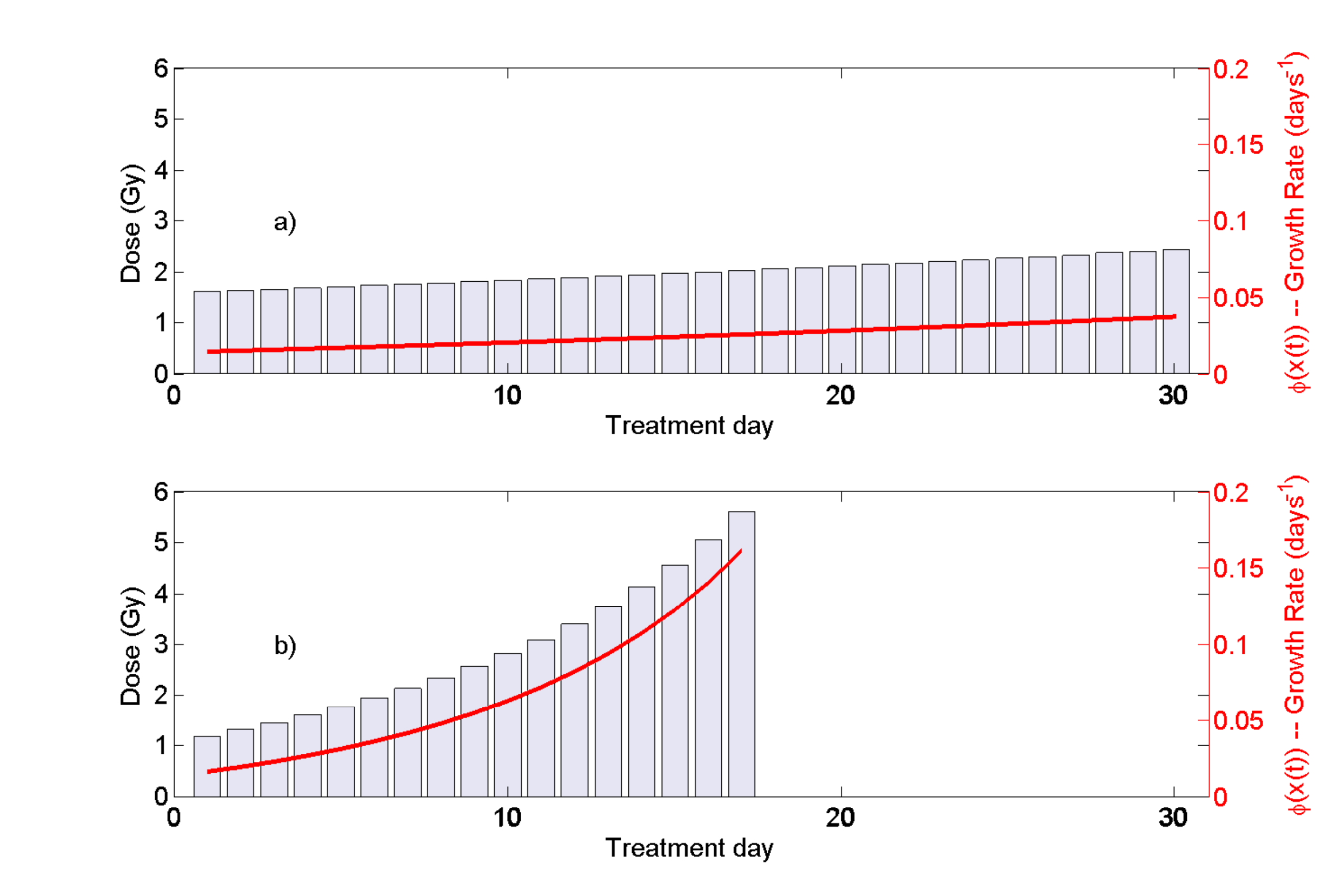}}
\caption{Optimal fractionation for accelerated repopulation in the case of $\abt = 5.7$ Gy. Plot a) shows the optimal fractionation schedule for a slowly proliferating tumor and plot b) for a fast one. The doubling time for the reference treatment begins at $\tau_d = 50$ days and decreases approximately to a) $20$ and b) $5$ days, respectively.}
\label{fig:plot2}
\end{figure}

We use a smaller value for the $\alpha/\beta$ value of the tumor and re-run the calculations from the previous subsection. The parameters of the Gompertzian growth remain the same for the slow and fast proliferating tumor. As before, we use the parameters $\abo = 3$ Gy, $\at = 0.3$ Gy$^{-1}$, $c=61.6$ Gy, and $\g = 0.7$, with the only change being that $\abt = 5.7$ Gy. Note that the condition $\abo = 3 < \g \abt = 4$ is satisfied, meaning that hypo-fractionation is not optimal (see Theorem \ref{thm3}). We run the DP algorithm for $N=1,2,\ldots,100$, and find $N=42$ and $N=17$ result in the best cost for the slowly and fast proliferating tumors, respectively. However, for practical purposes, we again set the maximum number of fractions to be $30$ for the slowly proliferating tumor because a very small fraction of the tumor's cells remain regardless of whether $N=42$ or $N=30$, and there is only a change of $0.7 \%$ in $Y_{N-1}^+$. As seen in part b) of Figure \ref{fig:plot2}, for the fast proliferating tumor, the sequence of radiation doses range from approximately $1$ Gy to $5.5$ Gy for the $17$ days of treatment. This change in the fractionation schedule results in the objective $Y_{N-1}^+$ to be $15.42$, as compared to $17.78$ for the reference treatment of $2$ Gy per day for $30$ days. This is a significant change of $13.3 \%$ in $Y_{N-1}^+$ and about $50.7 \%$ change in $X_{N-1}^+$ in comparison to the reference treatment. Similar to Section \ref{subsec:accel}, the fraction doses are approximately proportional to the instantaneous proliferation rate.

We can infer that a smaller $\alpha/\beta$ value of the tumor suggests using larger changes in the fraction sizes and shorter overall treatment duration; this results in larger gains in the objective value and hence in overall tumor control. Low values of $\alpha/\beta$ have been observed for disease sites such as prostate cancer (\cite{MRZ2012}). Numerical experiments also indicate a similar effect when varying the normal tissue sparing factor $\gamma$. That is, a smaller $\gamma$ results in larger changes in the fractionation schedule. Intuitively, this is because better sparing of normal tissue allows a more aggressive treatment with higher tumor control. Of course, if the $\alpha/\beta$ value or sparing factor $\gamma$ is very small, hypo-fractionation would be optimal (see Theorem \ref{thm3}).

\subsection{Effect of weekend breaks} \label{subsec:weekend}

\begin{figure}
\centerline{\includegraphics[width=160mm]{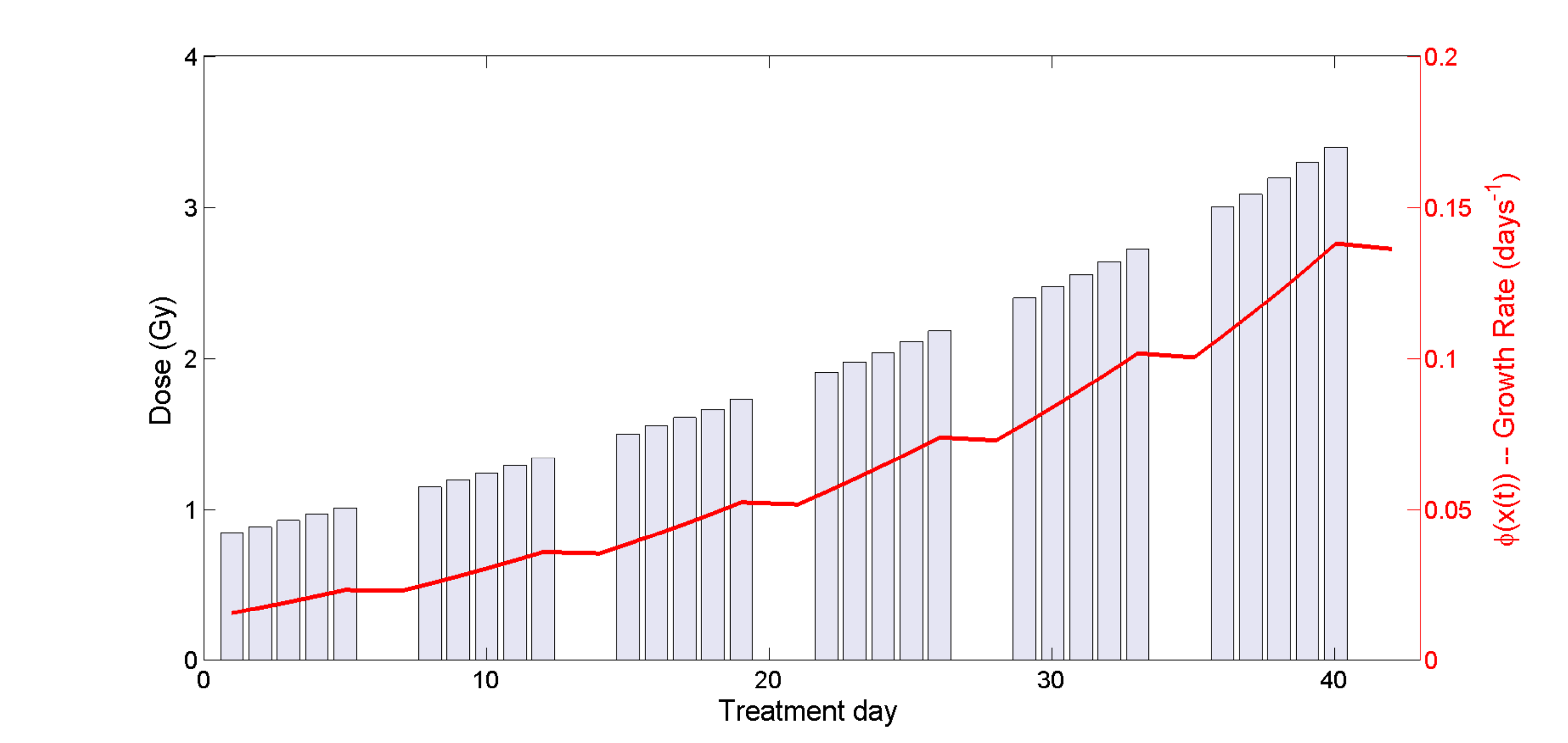}}
\caption{Effect of weekend breaks on optimal fractionation. We use $30$ treatment sessions and $\abt = 10$ Gy. The doubling time for the reference treatment begins at $\tau_d = 50$ days and decreases approximately to $5$ days.}
\label{fig:weekend}
\end{figure}

We simulate the effect of weekend breaks by using $30$ treatment sessions (starting Monday and ending Friday) with appropriately inserted weekend days, resulting in $N=40$ days ($6$ weeks, $5$ weekends). We set $\abt=10$ and assume a fast proliferating tumor, with Gompertzian growth parameters given in Subsection \ref{subsec:accel}. Thus, the parameters are the same as those used in part b) of Figure \ref{fig:plot1}, with the only difference being the inclusion of weekend breaks. The optimal fractionation scheme in this case is shown in Figure \ref{fig:weekend}. Notice that the dose fractions range from approximately 0.9 Gy in the first fraction to 3.5 Gy in the last fraction. The dose increase towards the end of treatment is larger compared to Figure \ref{fig:plot1} part b), which can probably be attributed to the lengthened overall treatment time from weekend breaks. Note that $\phi(x)$ is slightly decreasing during the weekend interval; this is because a growing tumor results in a smaller proliferation rate. In this numerical example, the decrease in $\phi(x)$ is small so that the dose fractions remain approximately proportional to $\phi(x)$. 

\section{Discussion and further remarks} \label{sec:disc}

\subsection{Non-uniform irradiation of the OAR} \label{subsec:nonuniform}

Although we have assumed throughout the paper that a dose $d$ results in a homogeneous dose $\g d$ to the OAR, in reality the OAR receives non-uniform irradiation. The tumor, however, is generally treated homogeneously. In \cite{UCS2013} and \cite{KHM2013}, the basic result stated in Theorem \ref{thm1} is generalized to arbitrary inhomogeneous doses in the OAR. The arguments in \cite{UCS2013} are also applicable to the case of repopulation as considered in this paper. We can define an effective sparing factor $\g_\text{eff}$ and an effective upper limit $c_\text{eff}$ on the $\text{BED}_O$, and the results in this paper will still hold. 

The results differ for the case of parallel OAR and serial OAR. A parallel organ could remain functional even with damaged parts; a serial OAR on the other hand remains functional only when all of its parts remain functional. For the case of a parallel OAR (e.g., lung), assuming that $\g_i d$ represents the dose in the $i$th voxel (or spatial point) in the OAR, the integral BED in the OAR is given by 
\begin{equation*}
\text{BED}_O = \sum_{k=0}^{N-1} \sum_i \g_i \, d_k \left(1+\frac{\g_i \, d_k}{\abo}\right).
\end{equation*}
After some algebraic manipulations, we obtain the same form for the normal tissue constraint as in this paper:
\begin{equation*}
\sum_{k=0}^{N-1} \g_\text{eff} \, d_k \left(1+\frac{\g_\text{eff} \, d_k}{\abo}\right) = c_\text{eff},
\end{equation*}
where $\g_\text{eff} = \sum_i \g_i^2 / \sum_i \g_i$ and $c_\text{eff} = c \g_\text{eff} / \sum_i \g_i$. For the serial case (e.g., spinal cord), only the maximum dose to the OAR matters, resulting in $\g_\text{eff} = \max_i \g_i$ and $c_\text{eff} = c$. Further details can be found in \cite{UCS2013}. When the OAR is neither completely parallel or completely serial, a good approximation of the BED may, for example, be a weighted combination of the BED for the parallel and serial cases. 

\subsection{Evolution of instantaneous proliferation rate}

In this work, we modeled accelerated repopulation as Gompertzian growth. For the numerical examples provided in Section \ref{num}, this leads to steadily increasing instantaneous proliferation rates over the course of treatment (see Figures \ref{fig:plot1} and \ref{fig:weekend}). For a smaller $\alpha/\beta$ value and a fast proliferating tumor, we even observed rapid increases in the instantaneous proliferation rate---see part b) of Figure \ref{fig:plot2}. However, more work is needed to estimate the temporal evolution of the repopulation rate $\phi(x)$ during treatment for specific disease sites.  

There are alternative tumor growth models such as the logistic and the Gomp-ex curve (\cite{Whe1988}) that we did not consider. However, the framework in this paper can also model these patterns of tumor growth.

There are some studies that indicate accelerated repopulation begins only after a lag period, e.g., for head and neck cancer (\cite{Bre1993}). We did not include such a lag period in our numerical experiments. But it can be modeled in our framework by using an appropriate form for $\phi(x)$. We expect a constant dose per fraction during this initial period of constant proliferation rate. 

\subsection{Variable time intervals}

The models in this paper are applicable for variable time intervals (e.g., weekends, holidays) as long as each radiation treatment is delivered in a short period of time, and the time interval between doses is comparatively long. However, our formulation is not applicable for very short time intervals (on the order of hours) between doses; in such cases, we would need to include biological effects such as incomplete repair of sublethal damage. For optimal fractionation schemes resulting from including these sublethal damage repair effects, see \cite{BBP2013}.

\subsection{Early and late responding normal tissue, and multiple OARs} \label{subsec:tissue} 

In the context of fractionation, two types of tissue are typically distinguished in the literature: early responding and late responding. Tumor tissue and some types of healthy tissue such as the skin are typically early responding. That is, they have a relatively large $\alpha/\beta$ value (e.g., 10), and they start proliferating within a few weeks of the start of radiation treatment (\cite{HaG2006}). On the other hand, late responding tissue typically have a low $\alpha/\beta$ value. In this paper, we considered a BED constraint for a single OAR, and we focused on the situation where $\abo < \g \abt$. This describes the situation in which a single late responding tissue is dose-limiting and needs to be spared via fractionation. This situation is expected to be most common. 

For the case where $\abo \geq \g \abt$, our model suggests a hypo-fractionation regimen. This condition can be fulfilled for early responding tissues which may exhibit a large enough $\abo$ value. However, this case requires careful consideration and possibly an extension of the model. For early responding tissue, the complication probability depends not only on the dose per fraction (as described via the $\abo$ value) but also the overall treatment time. This is currently not explicitly included in our model. 

In this paper, we have considered the case where a single OAR is dose-limiting. This is reasonable for some disease sites (e.g., in prostate cancer, the rectum is a single dose-limiting OAR). In the situation where multiple OARs are dose-limiting, additional BED constraints and therefore more states in the DP model would be required. This could be computationally intensive even with a few OARs; future research may address this situation. See also \cite{SGK2014,SGK2015} for some recent studies on optimal fractionation with multiple OARs.

\subsection{Comparison with prior work}

There have been studies that suggest dose escalation uniformly in time, in order to counter accelerated repopulation (see, for example, \cite{WaL2005} for prostate cancer and \cite{HMG2012} for cervical cancer). However, our paper primarily suggests dose intensification over time to counter the increased repopulation towards the end of treatment. Dose intensification has now been recommended by many studies, though in other contexts. The Norton-Simon hypothesis (\cite{NoS1977,NoS1986}) suggests increasing the dose intensity over the course of chemotherapy due to a Gompertzian tumor growth assumption. This work however dealt with chemotherapy and thus did not make use of the LQ model. Other studies suggest dose intensification to capitalize on the increased sensitivity of the tumor to radiation towards the end of treatment (\cite{HeW1973,AlB1976,MaD1991,WCW2000}). In \cite{MaD1991} and \cite{WCW2000}, a spherical tumor model consisting of a hypoxic core and an outer rim of well-nourished cells was used to analyze alternative fractionation schemes. The effect of radiation response, e.g., tumor shrinkage, here led to accelerated growth of the tumor towards the end of treatment. Our approach differs from these studies in that accelerated repopulation is modeled by letting the rate of repopulation depend directly on the number of cells in the tumor volume. A few studies indicate the effectiveness of concomitant boost therapy for head and neck cancers (\cite{Har1992,KiP1992}), where increased radiation is delivered at the end of treatment. These results seem to be consistent with the analysis in this paper. 


The dose intensification strategy presented in this paper has the potential to improve treatment outcomes for certain disease sites that exhibit significant accelerated repopulation, as long as the objective in our optimization model represents to some degree the actual objectives of the treatment. We note that current practice may actually result in the opposite of the suggested intensification strategy. The initial uniform doses prescribed are sometimes changed to reduced doses at the end of treatment to mitigate patient side effects. Intuitively, however, it is more beneficial to wait until the tumor is aggressive at the end of treatment and only then deliver a high dose, which may result in reduced side effects experienced at that time. Further studies investigating the clinical benefit of dose intensification to counter accelerated repopulation for specific disease sites would be useful. 

With technological advances such as functional and molecular imaging, there is potential to track previously unobservable biological processes such as the tumor proliferation rate during the course of therapy (\cite{BaS2008}). The opportunity to then adapt the treatment to the observed data, rather relying on a model, becomes a possibility. While previous works have investigated using imaging technology to select dose fractions (\cite{LCC2008,CLC2008,KGP2009,Kim2010,Gha2011,RCB2012}), they have not done so to counter accelerated repopulation. While the problem in this paper can be seen as a special case of the general one presented in \cite{KGP2012}, the effect of tumor proliferation, which is our main focus, was not analyzed therein. Prior works have also considered optimizing the number of treatment days \cite{Kim2010}, but again, have not done so for general tumor repopulation characteristics, including the case of accelerated repopulation.

The insights from this paper suggest that the increases in doses are approximately proportional to the proliferation rate, $\phi(x)$. This suggests the importance of further advancement of biological imaging technologies that can accurately measure quantities such as tumor proliferation rates during the course of treatment. An interesting approach worth investigating would then be the use of such imaging techniques to guide therapy. 

\section{Conclusions} \label{sec:con}

There are multiple ways to model accelerated repopulation. One approach could be to increase the tumor proliferation rate with already delivered dose or BED. In this paper, we chose instead to model accelerated repopulation implicitly by using a decelerating tumor growth curve, e.g., Gompertzian growth, with a proliferation rate $\phi(x)$ that is dependent on the number of tumor cells. This resulted in accelerated growth towards the end of the treatment due to fewer cells remaining after initial radiation treatment. We developed a DP framework to solve the optimal fractionation problem with repopulation for general tumor growth characteristics described by $\phi(x)$. We proved that the optimal dose fractions are non-decreasing over time, and showed the optimal number of fractions is finite. We derived the special structure of the problem when assuming Gompertzian tumor growth. This resulted in maximizing a discounted version of $\text{BED}_T$, which placed a higher weight on later treatment days, due to increased tumor proliferation. In this paper, we arrived at three main conclusions:
\begin{itemize}
\item Faster tumor growth suggested shorter overall treatment duration.
\item Accelerated repopulation suggested larger dose fractions later in the treatment to compensate for the increased tumor proliferation. Numerical results indicated that the optimal fraction sizes were approximately proportional to the instantaneous proliferation rate.
\item The optimal fractionation scheme used more aggressive increases in dose fractions with a shorter overall treatment duration when the $\alpha/\beta$ value of the tumor was smaller; in this case, there were larger gains in tumor control.
\end{itemize}

The advantage of the methods presented in this paper is that a change in the fractionation schedule can be readily implementable in a clinical setting, without technological barriers. However, the results presented in this paper are for illustrative purposes and should not be taken as immediate recommendations for a change in clinical practice. Clinical trials that compare standard approaches with intensified dose at the end of treatment would be needed to quantify the benefit for specific disease sites. We also realize that actual tumor dynamics are more complex than presented in this paper. The tumor volume may consist of a heterogeneous set of cells each with varying division rates. Effects such as re-oxygenation and incomplete repair have not been taken into account. Yet, we have avoided incorporating all of these aspects in a single model in order to primarily focus on the effect of accelerated repopulation. We hope that this analysis can provide useful insights and a basis for further research. 

%
\section{Appendix}

\subsection{Proof of Theorem \ref{thm1}} \label{sec:thm1}
\begin{proof}
A geometric proof is given in \cite{MTD2012}. An alternative proof is provided below. In the absence of repopulation, the fractionation problem simplifies to (\ref{opt_prob6}) without the second term in the objective, $(N-1) \rho / \at$. Thus, we would like to maximize $\text{BED}_T$ subject to $\text{BED}_O \leq c$. From Lemma \ref{lem3}, it is clear that the constraint on $\text{BED}_O$ is binding. We can now equivalently write the objective as 
\begin{equation*}
\text{BED}_T = \left( 1-\frac{\abo}{\g \abt} \right) \sum_{k=0}^{N-1} d_k + \frac{\abo}{\g^2 \abt} c. \label{ln1}
\end{equation*}
If $\abo = \g \abt$, then any feasible solution is optimal. If $\abo > \g \abt$, we would like to minimize $\sumD$ subject to a fixed BED$_O$. Otherwise, if $\abo < \g \abt$, we would like to maximize $\sumD$. First, let us assume the former condition $\abo > \g \abt$ to be true. Suppose an optimal solution is to deliver a non-zero dose in more than one fraction, i.e., that there exists $i$ and $j$ with $i \neq j$ such that $d_i^* > 0$ and $d_j^* > 0$. We would like to show that, leaving $d_k$ unchanged for $k \neq i,j$, an equally good or better solution is to set $d_i$ equal to $0$ and $d_j$ equal to the dose that would satisfy the constraint on $\text{BED}_O$ with equality. Essentially, this would mean that there is no reason to ``spread out" the dose among fractions; it would be optimal to deliver all the dose in a single fraction. We now set $d_i=0$ and $d_j=d_j^*+\delta$, where $\delta$ is a positive constant appropriately chosen so that the constraint on $\text{BED}_O$ is satisfied. Writing out the constraint equation and carrying out some algebra, we find
\begin{align*}
\sum_{k=0}^{N-1} \gamma d_k \left(1 + \frac{\gamma d_k}{[\alpha / \beta ]_O} \right) &= \sum_{\substack{k\neq i \\ k\neq j}} \gamma d_k^* \left(1 + \frac{\gamma d_k^*}{[\alpha / \beta ]_O} \right) + \gamma (d_j^*+\delta) \left(1 + \frac{\gamma (d_j^*+\delta)}{[\alpha / \beta ]_O} \right) \\
&= c + f_1(d_i^*, d_j^*, \delta), 
\end{align*}
where for the last equality we used the fact that the assumed optimal solution satisfies the constraint on $\text{BED}_O$ with equality. The function $f_1$ is defined as
\begin{equation}
f_1(d_i^*, d_j^*, \delta) = \gamma \delta \left(1 + \frac{\gamma \delta}{[\alpha / \beta ]_O} \right) - \gamma d_i^* \left(1 + \frac{\gamma d_i^*}{[\alpha / \beta ]_O} \right) + \frac{2 \g^2 \delta d_j^*}{[\alpha / \beta ]_O}. \label{funcf1}
\end{equation}
Thus, we can choose $\delta > 0$ that satisfies $f(d_i^*, d_j^*, \delta) = 0$ to ensure that the constraint on $\text{BED}_O$ is binding. Since the third term on the right hand side of (\ref{funcf1}) is positive, such a choice of $\delta$ will result in $\delta < d_i^*$. However, this means that we have found a strictly better solution because
\begin{equation*}
\sumD = \sum_{k\neq i} d_k^* + \delta < \sum_{k=0}^{N-1} d_k^*,
\end{equation*}
resulting in a contradiction of the optimality assumption. Therefore, for $\abo > \g \abt$, an optimal solution has exactly one non-zero dose. The cost from delivering this non-zero dose is the same regardless of which day it is delivered. The closed-form solution, as given in (\ref{mizuta1}), can be obtained by solving the quadratic equality constraint $\gamma d_j \left(1 + \frac{\gamma d_j}{[\alpha / \beta ]_O} \right) = c$.

Now, assume that $\abo < \g \abt$, and suppose that there exist $i$ and $j$ with $i \neq j$ such that $d_i^* \neq d_j^*$, with at least one of $d_i^*$ and $d_j^*$ non-zero. We would like to show that, leaving unchanged dose fractions at times other than $i$ or $j$, a better solution is to deliver equal dose in fractions $i$ and $j$, scaled appropriately so that the constraint on $\text{BED}_O$ is satisfied. Hence, we set $d_i=d_j=\delta(d_i^*+d_j^*)/2$, with an appropriately chosen positive constant $\delta$. Writing out the constraint equation and carrying out the algebra, we find
\begin{align}
\sum_{k=0}^{N-1} \gamma d_k \left(1 + \frac{\gamma d_k}{[\alpha / \beta ]_O} \right) &= \sum_{k \neq i, j} \gamma d_k^* \left(1 + \frac{\gamma d_k^*}{[\alpha / \beta ]_O} \right) + 2 \gamma \delta \frac{d_i^*+d_j^*}{2} \left(1 + \frac{\gamma \delta}{[\alpha / \beta ]_O} \frac{d_i^*+d_j^*}{2} \right) \nonumber \\
&= \sum_{k \neq i, j} \gamma d_k^* \left(1 + \frac{\gamma d_k^*}{[\alpha / \beta ]_O} \right) + f_2(d_i^*, d_j^*, \delta), \label{ln2}
\end{align}
where
\begin{equation*}
f_2(d_i^*, d_j^*, \delta) = \delta \g d_i^* \left(1 + \frac{\delta \g d_i^*}{[\alpha / \beta ]_O} \right) + \delta \g d_j^* \left(1 + \frac{\delta \g d_j^*}{[\alpha / \beta ]_O} \right) - \frac{(\g \delta)^2}{2 [\alpha / \beta ]_O} (d_i^* - d_j^*)^2. \label{funcf2}
\end{equation*}
Note that if we choose $\delta=1$, Equation (\ref{ln2}) would simplify to $c$ minus the third term in the definition of the function $f_2$. Clearly, in order to ensure that the expression in (\ref{ln2}) equals $c$ and thus satisfies the constraint on $\text{BED}_O$, $\delta > 1$ must be true. However, this means that we have found a strictly better solution because 
\begin{equation*}
\sumD = \sum_{\substack{k\neq i \\ k\neq j}} d_k^* + \delta (d_i^* + d_j^*) > \sum_{k=1}^N d_k^*,
\end{equation*}
resulting in a contradiction of the optimality assumption. Therefore, for $\abo < \g \abt$, the unique optimal solution consists of uniform doses. The closed-form solution, as given in (\ref{mizuta2}), can be obtained by solving the quadratic equality constraint $\gamma N d_j \left(1 + \frac{\gamma d_j}{[\alpha / \beta ]_O} \right) = c$. 
\end{proof}

\subsection{Proof of Theorem \ref{thm2}} 
\label{sec:thm2}
\begin{proof}
For a fixed $N$, we first determine the optimal dose fractions expressed as function of $N$ resulting from solving the problem (\ref{opt_prob6}). These optimal dose fractions are already given in Theorem \ref{thm1}. Now, we want to solve the optimization problem
\begin{equation}
\underset{N \in \mathbb{N}}{\text{maximize}} \quad -Y_0 - \frac{1}{\alpha_T} (N-1)\rho + \sum_{k=0}^{N-1} d_k^* \left( 1+\frac{d_k^*}{\abt} \right), \label{num_frac}
\end{equation}
where $d_k^*$ for $k=0,1,\ldots,N-1$ is given in Theorem \ref{thm1}. If $\abo \geq \g \abt$, from Theorem \ref{thm1}, it is optimal to deliver a single dose on one day. Thus, the last term in (\ref{num_frac}) is independent of $N$, and maximizing the second term results in $N^*=1$. For the case when $\abo < \g \abt$, we first show that the objective in (\ref{num_frac}) is strictly concave and eventually decreases as a function of $N$. Using the expression for $d_j^*$ given in Theorem \ref{thm1} and carrying out basic algebra, we find that the last term in (\ref{num_frac}) representing the BED in the tumor satisfies
\begin{align*}
\text{BED}_T &= \sum_{k=0}^{N-1} d_k^* \left( 1+\frac{d_k^*}{\abt} \right) \\
&= \frac{B}{\at} \sqrt{N^2+2AN} - \frac{B}{\at} N + E,
\end{align*}
where $A = \frac{2 c}{\abo}$, $B = \frac{\alpha_T \abo}{2\g} \left( 1 - \frac{\abo}{\g \abt} \right)$, and $E=\frac{\abo c}{\g^2 \abt}$. We want to show that the objective (\ref{num_frac}) is concave in $N$ for any $N \in [0,\infty)$. The domain $[0,\infty)$ is convex as desired. Now, we compute the first and second derivative with respect to $N$ of this objective. After some algebra, we obtain
\begin{equation}
\frac{\text{d}}{\text{d}N} \left[ -Y_0 - \frac{1}{\alpha_T}(N-1)\rho + \text{BED}_T \right] = \frac{B}{\at} \left( \frac{N+A}{\sqrt{(N+A)^2 - A^2}} - 1 \right) - \frac{\rho}{\at} \label{first_deriv}
\end{equation}
and
\begin{equation*}
\frac{\text{d}^2}{\text{d}^2N} \left[ -Y_0 - \frac{1}{\alpha_T}(N-1)\rho + \text{BED}_T \right] = - \frac{A^2 B}{\at \left(N^2 + 2AN \right)^{3/2}}.
\end{equation*}
Note that $A>0$, and since $\abo < \g \abt$, $B>0$ is also true. Thus, the second derivative above is strictly negative for all $N \in [0,\infty)$; this implies that the objective is strictly concave. As  $N$ grows large, the first term in the right-hand side of (\ref{first_deriv}) approaches $0$. Hence, the first derivative eventually becomes strictly negative due to the repopulation term, which means that the objective is a decreasing function for large enough $N$. Maximizing a strictly concave function that eventually decreases results in a unique optimum over the interval $[0,\infty)$. Now, we set the derivative (\ref{first_deriv}) equal to $0$ and solve for $N$. The result is
\begin{equation*}
N_c = A \left( \sqrt{\frac{\left( \rho + B \right)^2}{\rho \left( \rho + 2B \right)}} - 1\right).
\end{equation*}
Due to the concavity property of the objective, the additional constraint that $N$ is a natural number means that the optimum $N^*$ is either $\lfloor N_c \rfloor$ or $\lceil N_c \rceil$, whichever results in a larger objective value. The only exception is when $N_c < 1$, in which case $N^* = 1$ since $N^*$ cannot be $0$. 
\end{proof}

\subsection{Proof of Lemma \ref{lem3}} 
\label{sec:lem3}
\begin{proof}
Let us suppose an optimal solution is $d_i^*$, for $i=0,1,\ldots,N-1$, but the constraint on $\text{BED}_O$ is not active. That is, there exists $\delta > 0$ such that 
\begin{equation*}
\sum_{k=0}^{N-1} \gamma d_k^* \left(1 + \frac{\gamma d_k^*}{[\alpha / \beta ]_O} \right) + \delta = c.
\end{equation*}
For every fraction $i$, let us increase $d_i^*$ by the same positive constant $\eps$. Choosing $\eps$ small enough so that $\sum_{k = 0}^{N-1} \g \eps \left( 1 + \frac{\g}{[\alpha / \beta]_O} \left( 2 d_k^* + \eps \right) \right) < \delta$, we see that
\begin{align*}
\sum_{k=0}^{N-1} \gamma (d_k^* +\eps) \left(1 + \frac{\gamma (d_k^*+\eps)}{[\alpha / \beta ]_O} \right) &=  \sum_{k=0}^{N-1} \gamma d_k^* \left(1 + \frac{\gamma d_k^*}{[\alpha / \beta ]_O} \right) + \sum_{k = 0}^{N-1} \g \eps \left( 1 + \frac{\g}{[\alpha / \beta]_O} \left( 2 d_k^* + \eps \right) \right) \\
&< \sum_{k=0}^{N-1} \gamma d_k^* \left(1 + \frac{\gamma d_k^*}{[\alpha / \beta ]_O} \right) + \delta \\
&= c,
\end{align*}
which means that the constraint on $\text{BED}_O$ is still satisfied. Thus, for every $i$, we are able to deliver $d_i = d_i^*+\eps$ while ensuring the constraint on $\text{BED}_O$. Let the sequence $\{ \widetilde{Y}_i^+ \}$ and the sequence $\{ Y_i^+ \}$ be the result of respectively delivering $d_i^*$ and $d_i$, for $i=0,1,\ldots, N-1$. We will show by induction that $Y_i^+ < \widetilde{Y}_i^+$ for every $i$, which would result in a contradiction of the optimality assumption. For the base case, we have
\begin{equation*}
Y_0^+ = Y_0 - \text{BED}_T(d_0) < Y_0 - \text{BED}_T(d_0^*) = \widetilde{Y}_0^+
\end{equation*}
because $d_0 > d_0^*$. For the inductive step, suppose $Y_i^+ < \widetilde{Y}_i^+$. Then, we have
\begin{align*}
Y_{i+1}^+ &= F(Y_{i}^+) - \text{BED}_T(d_{i+1}) \\
&< F(\widetilde{Y}_i^+) - \text{BED}_T(d_{i+1}) \\
&< F(\widetilde{Y}_i^+) - \text{BED}_T(d_{i+1}^*) \\
&= \widetilde{Y}_{i+1}^+,
\end{align*}
where the first inequality holds because the growth function $F(\cdot)$ is strictly increasing due to the assumption $\phi(x)>0$, and the second inequality holds because $d_{i+1} > d_{i+1}^*$. The induction is complete, and we have shown $Y_{N-1}^+ < \widetilde{Y}_{N-1}^+$. This is a contradiction of the optimality assumption because we have found a strictly better solution. 
\end{proof}

\subsection{Proof of Lemma \ref{lem2}} 
\label{sec:lem2}
\begin{proof}
Since $\phi(x)>0$ for all $x$, the rate of change of the number of tumor cells is positive. Thus, assuming no delivered radiation, the number of cells is strictly increasing due to repopulation. Now, suppose $\yip<\widetilde{Y}_{i}^+$. It is clear that $Y_{i+1}^- < \widetilde{Y}_{i+1}^-$. Since $\phi(x)$ is non-increasing in $x$, we also have $\widetilde{Y}_{i+1}^- - Y_{i+1}^- \leq \widetilde{Y}_{i}^+ - Y_{i}^+$. And, after applying $d_{i+1}$, we also have $Y_{i+1}^+ < \widetilde{Y}_{i+1}^+$ and $\widetilde{Y}_{i+1}^+ - Y_{i+1}^+ \leq \widetilde{Y}_{i}^+ - Y_{i}^+$. The lemma follows by induction.
\end{proof}

\subsection{Proof of Theorem \ref{thm3}} \label{sec:thm3}
\begin{proof}
Suppose that an optimal solution to the fractionation problem with repopulation is the sequence of doses $d_0^*,d_1^*,\ldots,d_{N-1}^*$, resulting in the sequence $\widetilde{Y}_{0}^+,\widetilde{Y}_{1}^+,\ldots,\widetilde{Y}_{N-1}^+$. We will prove the results in this theorem by contradiction and finding another sequence $d_0,d_1,\ldots,d_{N-1}$, resulting in $Y_{0}^+,Y_{1}^+,\ldots,Y_{N-1}^+$, that has strictly better cost. 

Fix some $i$ and $j$, with $i < j$, and let $d_k=d_k^*$ for $k\neq i,j$. We will first show that if $Y_{i}^+>\widetilde{Y}_{i}^+$ (or, equivalently, $d_i < d_i^*$), then $Y_{j}^+ \leq \widetilde{Y}_{j}^+ + \text{BED}_T^* - \text{BED}_T$, where $\text{BED}_T^*$ and $\text{BED}_T$ are the total BED in the tumor resulting from delivering $\{d_k^*\}$ and $\{d_k\}$, respectively. Suppose $Y_{i}^+>\widetilde{Y}_{i}^+$. Then, clearly $\text{BED}(d_i^*)>\text{BED}(d_i)$. Before delivering the dose $d_i$, the same sequences of doses were applied; this means that $Y_i^- = \widetilde{Y}_i^-$. Now, $Y_i^+  = Y_i^- - \text{BED}_T(d_i)$ and $\widetilde{Y}_i^+  = \widetilde{Y}_i^- - \text{BED}_T(d_i^*)$, implying that $Y_i^+ - \widetilde{Y}_i^+ = \text{BED}_T(d_i^*) - \text{BED}_T(d_i)$. Since $\phi(x)$ is non-increasing as a function of $x$, from Lemma \ref{lem2}, we have that 
\begin{equation}
Y_{j}^- - \widetilde{Y}_{j}^- \leq \text{BED}_T(d_i^*) - \text{BED}_T(d_i). \label{ineq_bound}
\end{equation}
Now, we have
\begin{align}
\yjp &= \yjm - \text{BED}_T(d_j) \nonumber \\
&= \yjm - \text{BED}_T(d_j) + \widetilde{Y}_j^+ - (\widetilde{Y}_j^- - \text{BED}_T(d_j^*)) \nonumber \\
&= \widetilde{Y}_j^+ + (\yjm - \widetilde{Y}_j^-) + \text{BED}_T(d_j^*) - \text{BED}_T(d_j) \nonumber \\
&\leq \widetilde{Y}_j^+ + \text{BED}_T(d_i^*) - \text{BED}_T(d_i) + \text{BED}_T(d_j^*) - \text{BED}(d_j) \nonumber \\
&= \widetilde{Y}_j^+ + \text{BED}_T^* - \text{BED}_T, \label{ineq_bound2}
\end{align}
where the inequality is due to (\ref{ineq_bound}).

Suppose that $\abo > \g \abt$ and that there exists $i < N-1$ such that $d_i^*>0$. We set $d_i = 0$ and $d_{N-1} = d_{N-1}^* + \delta$, where $\delta$ is a positive constant that enforces the constraint on $\text{BED}_O$. Then, we find, using exactly the same argument as in the proof of Theorem \ref{thm1}, that $\text{BED}_T^* < \text{BED}_T$. Since $d_i=0$ and $d_i^*>0$ imply that $\yip > \widetilde{Y}_i^+$, by taking $j = N-1$ we have from (\ref{ineq_bound2}) that $Y_{N-1} ^+ < \widetilde{Y}_{N-1}^+$. We have found a strictly better solution than the supposed optimal one. Thus, if $\abo > \g \abt$, the optimal solution is to deliver a single dose on the last day of treatment. 

Suppose now $\abo < \g \abt$ and there exist $i$ and $j$, with $i < j$, such that $d_i^*>d_j^*$. Let $d_i = d_j = \delta \left(\frac{d_i^*+d_j^*}{2}\right)$ where $\delta$ is chosen so as to ensure the BED constraint in the OAR. Again, we use the same argument in Theorem \ref{thm1} and conclude that $\text{BED}_T^* < \text{BED}_T$. Since $d_i<d_i^*$, we have $\yip > \widetilde{Y}_i^+$. From (\ref{ineq_bound2}), we conclude $\yjp < \widetilde{Y}_j^+$. This means that $Y_{N-1}^+ < \widetilde{Y}_{N-1}^+$ due to Lemma \ref{lem2}. Thus, we have found a strictly better solution than the supposed optimal one. Therefore, when $\abo < \g \abt$, the optimal doses must increase over the course of treatment, i.e., $d_0^* \leq d_1^* \leq \cdots \leq d_{N-1}^*$. 
\end{proof}

\subsection{Proof of Theorem \ref{thm4}} \label{sec:thm4}
\begin{proof}
When $\abo \geq \g \abt$, we already know from Theorem \ref{thm3} that $N^*=1$ and thus the optimal number of fractions is finite. We now give the proof for the case where $\abo < \g \abt$. For a fixed $N$, we assume $d_0^*,d_1^*,\ldots,d_{N-1}^*$ is a sequence of optimal dose fractions resulting from minimizing the logarithm version of the expected number of cells $Y_{N-1}^+$ subject to the constraint on $\text{BED}_O$. From the equation $\text{d}y/\text{d}t=\phi(\exp(\at y(t)))/\at$, the assumption $\phi(x)>r$ implies $\text{d}y/\text{d}t > r'$  for some other constant $r'>0$. The function mapping $Y^+$ to $Y^-$ as a result of repopulation satisfies
\begin{equation*}
F(Y^+) > Y^+ + r'.
\end{equation*}
Now, since $F$ is applied $N-1$ times, we can bound $Y_{N-1}^+$ from below:
\begin{equation*}
Y_{N-1}^+ > Y_0 - \text{BED}_T^* + (N-1) r'
\end{equation*}
where $\text{BED}_T^*$ is the BED in the tumor resulting from delivering the doses $d_0^*,d_1^*,\ldots,d_{N-1}^*$. From Theorem \ref{thm1}, we know that when $\abo < \g \abt$, uniformly distributed doses maximize $\text{BED}_T$ subject to the constraint on $\text{BED}_O$. Let $\widetilde{\text{BED}}_T$ represent the BED in the tumor resulting from delivering these uniformly distributed doses. Then, $\text{BED}_T^* \leq \widetilde{\text{BED}}_T$ and we have
\begin{equation*}
Y_{N-1}^+ > Y_0 -\text{BED}_T^* + (N-1) r' \geq Y_0 -\widetilde{\text{BED}}_T + (N-1) r'.
\end{equation*}
Using the expression for $d_j^*$ given in Theorem \ref{thm1} and carrying out basic algebra as also done in Theorem \ref{thm2}, we find that
\begin{align*}
\widetilde{\text{BED}}_T &= \sum_{k=0}^{N-1} d_k^* \left( 1+\frac{d_k^*}{\abt} \right) \\
&= \frac{B}{\at} \sqrt{N^2+2AN} - \frac{B}{\at} N + E,
\end{align*}
where $A = \frac{2 c}{\abo}$, $B = \frac{\alpha_T \abo}{2\g} \left( 1 - \frac{\abo}{\g \abt} \right)$, and $E=\frac{\abo c}{\g^2 \abt}$. After some algebra, we obtain
\begin{equation*}
\frac{\text{d}}{\text{d}N} \left[Y_0 -\widetilde{\text{BED}}_T + (N-1)r' \right] = -\frac{B}{\at} \left( \frac{N+A}{\sqrt{(N+A)^2 - A^2}} - 1 \right) + r'.
\end{equation*}
Since the first derivative above eventually stays positive for large enough $N$, $Y_{N-1}^+$ approaches infinity as $N\rightarrow \infty$. Thus, the optimal number of dose fractions $N^*$ cannot be infinite.
\end{proof}

\subsection{Proof of Corollary \ref{coro1}} \label{sec:coro1}
\begin{proof}
When we include breaks and fix some dose fractions, we note that Lemma \ref{lem2} and its proof remain valid for all other fractions. The result follows using the same argument as in the proof of Theorem \ref{thm3}.
\end{proof}

%
%



\section{Acknowledgements}

The authors thank David Craft and Ehsan Salari for helpful discussions and feedback. This research was partially funded by Siemens and was performed while the 2nd author was with the Laboratory for Information and Decision Systems, Massachusetts Institute of Technology. Thanks to anonymous reviewers for pointing out a generalization of Corollary \ref{coro1} and also for very helpful feedback that has improved this paper.


\begin{spacing}{0.9}
\bibliographystyle{abbrv}
\bibliography{ref}
\end{spacing}

\end{document}